\documentclass[12pt]{article}
\usepackage[dvips]{graphicx}
\usepackage{amssymb}
\usepackage{amsmath}
\usepackage{here}
\usepackage{color}
\usepackage{subfigure}
\usepackage{cite}

\newcommand{\bear}{\begin{array}}  
\newcommand {\eear}{\end{array}}
\newcommand{\bea}{\begin{eqnarray}}   
\newcommand{\eea}{\end{eqnarray}}
\newcommand{\beq}{\begin{equation}}   
\newcommand{\eeq}{\end{equation}}
\newcommand{\bef}{\begin{figure}}  \newcommand 
{\eef}{\end{figure}}
\newcommand{\bec}{\begin{center}}  \newcommand 
{\eec}{\end{center}}

\renewcommand{\thefootnote}{\fnsymbol{footnote}}
\begin{document}

\begin{titlepage}

\begin{flushright}
ICRR-Report-647-2012-36\\
\end{flushright}

\vskip 2.0cm

\begin{center}

{\large \bf Forecast constraints on cosmic strings from future CMB, pulsar timing and gravitational wave direct detection experiments.  
}

\vskip 1.2cm

Sachiko Kuroyanagi$^a$
Koichi Miyamoto$^b$,
Toyokazu Sekiguchi$^{c}$,
Keitaro Takahashi$^{d}$\\
and Joseph Silk$^{e,f}$

\vskip 0.4cm

{\it $^a$Research Center for the Early Universe (RESCEU),
University of Tokyo, Tokyo 113-0033, Japan}\\
{\it $^b$Institute for Cosmic Ray Research,
University of Tokyo, Kashiwa 277-8582, Japan}\\
{\it $^c$Department of Physics and Astrophysics, Nagoya University, Nagoya 464-8602, Japan}\\
{ \it $^d$Faculty of Science, Kumamoto University, 2-39-1, Kurokami, Kumamoto 860-8555, Japan}\\
{\it $^e$Institut d' Astrophysique, 98bis Boulevard Arago, Paris 75014, France}\\
{\it $^f$Department of Physics, University of Oxford, Keble Road, Oxford, OX1 3RH, UK}

\vskip 1.2cm

\date{\today}

\begin{abstract}
  We study future observational constraints on cosmic string
  parameters from various types of next-generation experiments: direct
  detection of gravitational waves (GWs), pulsar timing array, and the
  cosmic microwave background (CMB).  We consider both GW burst and
  stochastic GW background searches by ground- and space-based
  interferometers as well as GW background detection in pulsar timing
  experiments.  We also consider cosmic string contributions to the
  CMB temperature and polarization anisotropies.  These different
  types of observations offer independent probes of cosmic strings and
  may enable us to investigate cosmic string properties if the
  signature is detected.  In this paper, we evaluate the power of
  future experiments to constrain cosmic string parameters, such as
  string tension $G\mu$, initial loop size $\alpha$, and 
  reconnection probability $p$, by performing Fisher information
  matrix calculations.  We find that combining the information from
  the different types of observations breaks parameter degeneracies
  and provides more stringent constraints on the parameters.  We also
  find future space-borne interferometers independently provide a
  highly precise determination of the parameters.
\end{abstract}

\end{center}
\end{titlepage}

\renewcommand{\thefootnote}{\arabic{footnote}}

\section{Introduction}

Cosmic strings are linear topological defects which are formed at
spontaneous symmetry breaking (SSB) in the early Universe
\cite{Kibble:1976sj} (as a review, see Ref. \cite{Vilenkin}).  Some
inflation models based on superstring theory predict fundamental
strings, D-strings and their bound states of cosmological length.
Such stringy cosmic strings are called cosmic superstrings
\cite{Sarangi:2002yt,Jones:2003da,Dvali:2003zj}.  They form a
complicated string network, which consists of infinite strings and
closed loops, and may leave remarkable signatures in the universe
through their nonlinear evolution.  If their signals are observed, not
only the existence of cosmic strings will be confirmed but their
properties might be studied.  This enables us to obtain implications
for both the history of the universe such as inflation or SSB, and for
physics beyond the Standard Model of particle physics such as grand
unified theory (GUT) or superstring theory through the study of cosmic
strings.  Therefore, cosmic strings are important probes of both
cosmology and particle physics, and we are motivated to study how the
properties of cosmic strings can be determined by future experiments.

Various types of observational signatures of comic strings have been
studied intensively.  One of them is the gravitational wave (GW)
\cite{Vilenkin:1981bx,Hogan:1984is,Caldwell:1991jj,Caldwell:1996en,Damour:2000wa,Damour:2001bk,Damour:2004kw,Siemens:2006yp,DePies:2007bm,Kawasaki:2010yi,Olmez:2010bi,Binetruy:2012ze}.
The main source of GWs in the string network is cusps on loops.
\footnote{ 
Concerning the stochastic GW background,
kinks make contribution comparable with cusps\cite{Olmez:2010bi},
which we do not take into account in this paper.
For cosmic superstrings each loop may have many kinks on itself,
and in such a case GWs from kinks may dominate that from cusps 
both in the rate of rare bursts and the amplitude of the stochastic background
 \cite{Binetruy:2010cc,Bohe:2011rk}. However, such a
  contribution strongly depends on the fraction of loops with
  junctions.  So we do not consider their contribution in this
  paper.   }
  A cusp is a highly Lorentz boosted region on a loop which
appears $\mathcal{O} (1)$ times in an oscillation period of the loop
and it emits a strong beam of GWs, which we call a ``GW burst''.  The
GW bursts can be detected in two different forms.  One is ``rare
bursts", which are infrequent but strong enough to be detected alone.
The other is a stochastic GW background, which consists of many small
bursts overlapping each other \cite{Damour:2000wa,Damour:2001bk}.  In
our previous paper \cite{Kuroyanagi:2012wm}, we studied how cosmic
string parameters are determined by direct detection of GWs in future
ground-based GW experiments, such as Advanced LIGO\cite{Sigg:2008zz},
Advanced Virgo\cite{Accadia:2011zzc} and KAGRA\cite{Kuroda:2010zzb},
and showed that measurements of the burst rate and the GW background
provide different information and leads to better constraints on
parameters when they are combined.

In this paper, we extend our previous study in two ways.  First, in
addition to the ground-based detectors, we consider other types of GW
experiment, that is, direct detection by space-borne interferometers
such as eLISA/NGO\cite{AmaroSeoane:2012km}, BBO\cite{BBO} or
DECIGO\cite{Kawamura:2011zz}, and observation of the GW background in
pulsar timing experiments, such as Parkes PTA\cite{Manchester:2007mx},
EPTA\cite{Ferdman:2010xq,vanHaasteren:2011ni}, NANOGrav\cite{Jenet:2009hk,Demorest:2012bv}
or SKA\cite{SKA}.  These different types of GW experiment provide
different information, since each type of experiment has its best
sensitivity at different frequency: $\sim 10^2$Hz for ground-based
detectors, $\sim 10^{-2}$Hz for eLISA/NGO, $\sim 10^{-1}$Hz for DECIGO
and BBO, and $\sim 10^{-8}$Hz for pulsar timing experiments.
Gravitational waves in different frequency bands are emitted at
different redshifts and carry information on cosmic strings living in
different epochs of the Universe.

Second, we also take into account observation of the cosmic microwave
background (CMB).  Cosmic strings may induce temperature and
polarization fluctuations in CMB through gravitational effects, which
is also an important observational signature of cosmic strings and
studied intensively (for example, see Refs.
\cite{Kaiser:1984iv,Gott:1984ef,Seljak:1997ii,Pogosian:1999np,Benabed:2000tr,Pogosian:2003mz,Seljak:2006hi,Bevis:2007gh,Bevis:2007qz,Pogosian:2007gi,Kawasaki:2010iy,Battye:2010xz,Bevis:2010gj,Yamauchi:2010ms,Mukherjee:2010ve}).
Although current CMB observations indicate that cosmic strings are not
the dominant source of CMB fluctuations, the string signatures might
be observed in small-scale fluctuations and/or B-mode polarization by
future experiments such as Planck\cite{Planck:2006aa} and
CMBpol\cite{Baumann:2008aq}.  Expected constraints on cosmic string
parameters by these experiments are already studied in
Ref. \cite{Foreman:2011uj} by Fisher information matrix calculations.
In this paper, we consider combination of the constraints with GW
experiments, which are expected to provide different information on
cosmic strings.

As in Ref.  \cite{Kuroyanagi:2012wm}, in this paper, we focus on three
parameters which characterize cosmic string network.  The first one is
string tension $\mu$, or the product of it and Newton constant $G$,
$G\mu$.  It represents the energy stored per unit length in a cosmic
string.  For the field theoretic string, it is comparable to the
square of the energy scale of SSB, while for the cosmic superstring,
it is determined by the energy scale of the superstring theory and the
warp factor of the extra dimension where the string is located.  The
value of $G\mu$ affects not only the amplitudes of the GW bursts and
GW background but also the spectral shapes through the change of the
lifetime of loops.  The amplitude of CMB fluctuations is also affected
by the value of $G\mu$.

The second one is initial loop size $\alpha$.  The typical size of a
loop at its formation is characterized by $\alpha t$, where $t$ is the
time of the loop formation.  In principle, the value of $\alpha$ can
be predicted, if we can solve the nonlinear evolution of the string
network, and there are many works which attempt to determine the value
of $\alpha$ with numerical or analytical methods
\cite{Bennett:1987vf,Allen:1990tv,Vincent:1996rb,Vanchurin:2005pa,Ringeval:2005kr,Martins:2005es,Olum:2006ix,BlancoPillado:2011dq,Siemens:2002dj,Polchinski:2006ee,Dubath:2007mf,Vanchurin:2007ee,Lorenz:2010sm}.
However, it is not yet clearly understood, so we treat $\alpha$ as a
free parameter.  The value of $\alpha$ affects both the GW burst rate
and the GW background spectrum.  On the other hand, the CMB signature
is independent from $\alpha$, since it is induced mainly by infinite
strings and the contribution from loops is negligible.

The third one is reconnection probability $p$.  For field theoretic
strings, $p$ is roughly equal to unity while for cosmic superstrings,
it can be much smaller than $1$.  The string network becomes denser as
$p$ gets smaller.  This leads to the enhancement of the burst rate,
the GW background spectrum, and the CMB fluctuations.  The shape of
the CMB power spectrum is also affected, since the typical length
scale of the string network becomes smaller and the average velocity
of the strings becomes larger.

In this paper, we aim to investigate the abilities of the different
types of experiments to constrain string parameters and study how they
complement each other.  We calculate the burst rate and the GW
background spectrum using the formulation described in Ref.
\cite{Kuroyanagi:2012wm}.  For CMB, we use CMBACT\cite{CMBACT}, which
calculates the power spectrum by approximating the string network as an
ensemble of randomly oriented straight segments\cite{Pogosian:1999np}.
Finally, we predict constraints on $G\mu$, $\alpha$ and $p$ from the
future experiments by performing Fisher matrix calculations and find
that the different types of experiment break the degeneracies in the
parameters and help to tighten the constraints when they are combined.

This paper is constructed as follows.  In Sec. 2, we briefly review
the model to describe the cosmic string network and the methods to
calculate GWs from string loops and CMB power spectra induced by
strings.  In Sec. 3, we describe the sensitivities of current and
future experiments and the Fisher matrix formalism for direct
detection, pulsar timing, and CMB experiments.  In Sec. 4, we first
show the parameter space to be explored by these experiments, and then
calculate constraints on the cosmic string parameters for three
fiducial models.  We summarize the paper in Sec. 5.  Throughout the
paper, we use the cosmological parameters from the 7-year WMAP data
(WMAP+BAO+H0 mean) \cite{Komatsu:2010fb}: the ratio of the present
energy density of baryon to the critical density
$\Omega_bh^2=0.02255$, that of cold dark matter $\Omega_ch^2=0.1126$,
that of dark energy $\Omega_\Lambda=0.725$, the spectral index of the
primordial curvature perturbation $n_s=0.968$, the reionization
optical depth $\tau=0.088$, the amplitude of the primordial curvature
perturbation $\Delta^2_{\mathcal{R}}(k_0)=2.43 \times 10^{-9}$ at
$k_0=0.002{\rm Mpc}^{-1}$, Hubble constant $H_0=70.2 {\rm km}/{\rm
  s}/{\rm Mpc}$ and the primordial helium abundance $Y_p=0.326$.

\section{Calculation of various types of cosmic string signatures}

In this section, we briefly mention how to calculate the observational
signatures of cosmic strings.  After explaining the analytic model of
the cosmic string network, we describe the formalism to calculate the
burst rate, the GW background spectrum and the CMB power spectrum and
mention their dependence on string parameters.

\subsection{The model of the cosmic string network}

As in our previous paper\cite{Kuroyanagi:2012wm}, we adopt the model
in Refs.  \cite{Avgoustidis:2005nv,Takahashi:2008ui}, which is based
on the velocity-dependent one-scale model \cite{Martins:1996jp}.  The
network of infinite strings is considered as a random walk with a
correlation length $\xi$, which corresponds to the typical curvature
radius and interval of infinite strings, and the total length $L$ of
infinite strings in volume $V$ is given by $L=V/\xi^2$.  The equations
for $\gamma\equiv\xi/t$ and the root mean square velocity of infinite
strings $v$ are given by
\beq
\frac{t}{\gamma}\frac{d\gamma}{dt}=-1+Ht +\frac{\tilde{c}(t)pv}{2\gamma}+Ht v^2,
\label{gammaeq}
\eeq 
\beq
\frac{dv}{dt}= (1-v^2)H \left(\frac{k (v)}{Ht\gamma}-2v\right),
\label{veq}
\eeq
where $k(v)=\frac{2\sqrt{2}}{\pi}\frac{1-8v^6}{1+8v^6}$
\cite{Martins:2000cs}.   
The Hubble parameter is given by 
\beq
H(t)\equiv \frac{\dot
  a}{a}=H_0\left[\Omega_{\Lambda}+\Omega_m(1+z)^3+\Omega_r(1+z)^4\right]^{1/2},
\eeq
where $a(t)$ is the scale factor, $1+z=a_0/a(t)$ is the redshift,
$a_0$ is the present value of the scale factor,
$\Omega_m=\Omega_b+\Omega_c$ and $\Omega_r$ is the ratio of the
present energy density of radiation to the critical density.  The
parameter $\tilde{c}(t)$ represents the efficiency of loop formation.
The value of this parameter between the radiation-dominated era and
the matter-dominated era is interpolated by 
\beq
\tilde{c}(t)=\frac{c_r+\frac{gc_m}{1+z}}{1+\frac{g}{1+z}}, 
\eeq 
where we set $c_r=0.23, c_m=0.18$ and $g=300$ according to \cite{Battye:1997hu,Pogosian:1999np}.
With these values, the evolution of gamma and v agrees with results from numerical simulations such as \cite{Martins:1996jp}.

The parameters $\gamma(t)$ and $v(t)$ have different time evolutions
depending on the Hubble expansion rate.  In our previous paper, we
used the asymptotic values for each radiation-dominated and
matter-dominated Universe.  However, in this paper, since we
additionally investigate the effect on the CMB fluctuations, which are
produced near the transition from the radiation-dominated to the
matter-dominated Universe, we numerically solve Eqs. (\ref{gammaeq})
and (\ref{veq}) to evaluate the values of $\gamma(t)$ and $v(t)$.
Their initial values are determined by the solution of $d\gamma/dt=0$
and $dv/dt=0$ in the radiation-dominated era with $\tilde c=c_r$ and
$Ht=1/2$.

\begin{figure}[t]
\begin{minipage}{1\hsize}
\begin{center}
\subfigure[$\gamma$ for $p=1,10^{-1},10^{-2}$.]{
\includegraphics[width=70mm]
{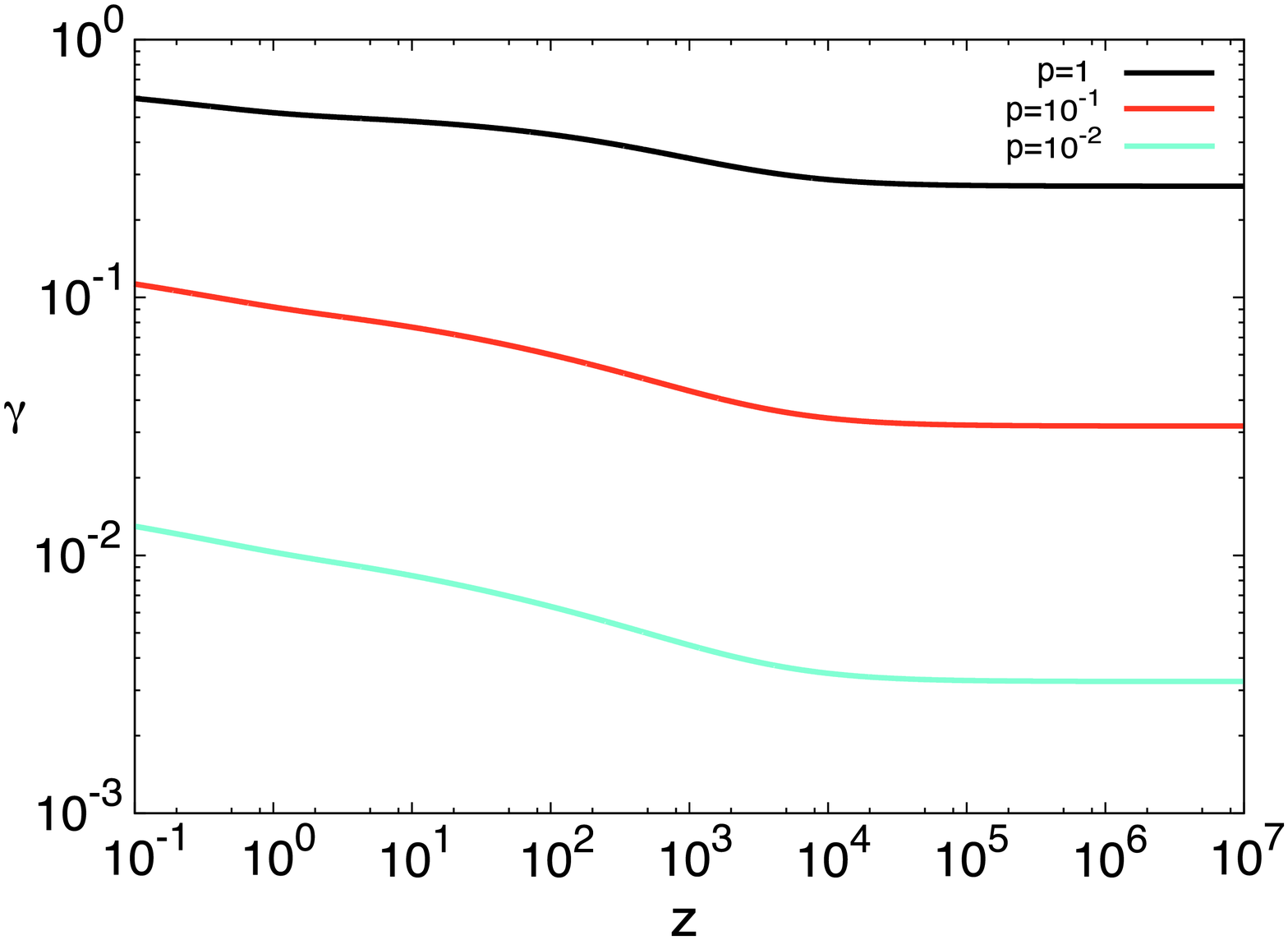}
\label{fig:gamma}
}
\subfigure[$v$ for $p=1,10^{-1},10^{-2}$.]{
\includegraphics[width=70mm]
{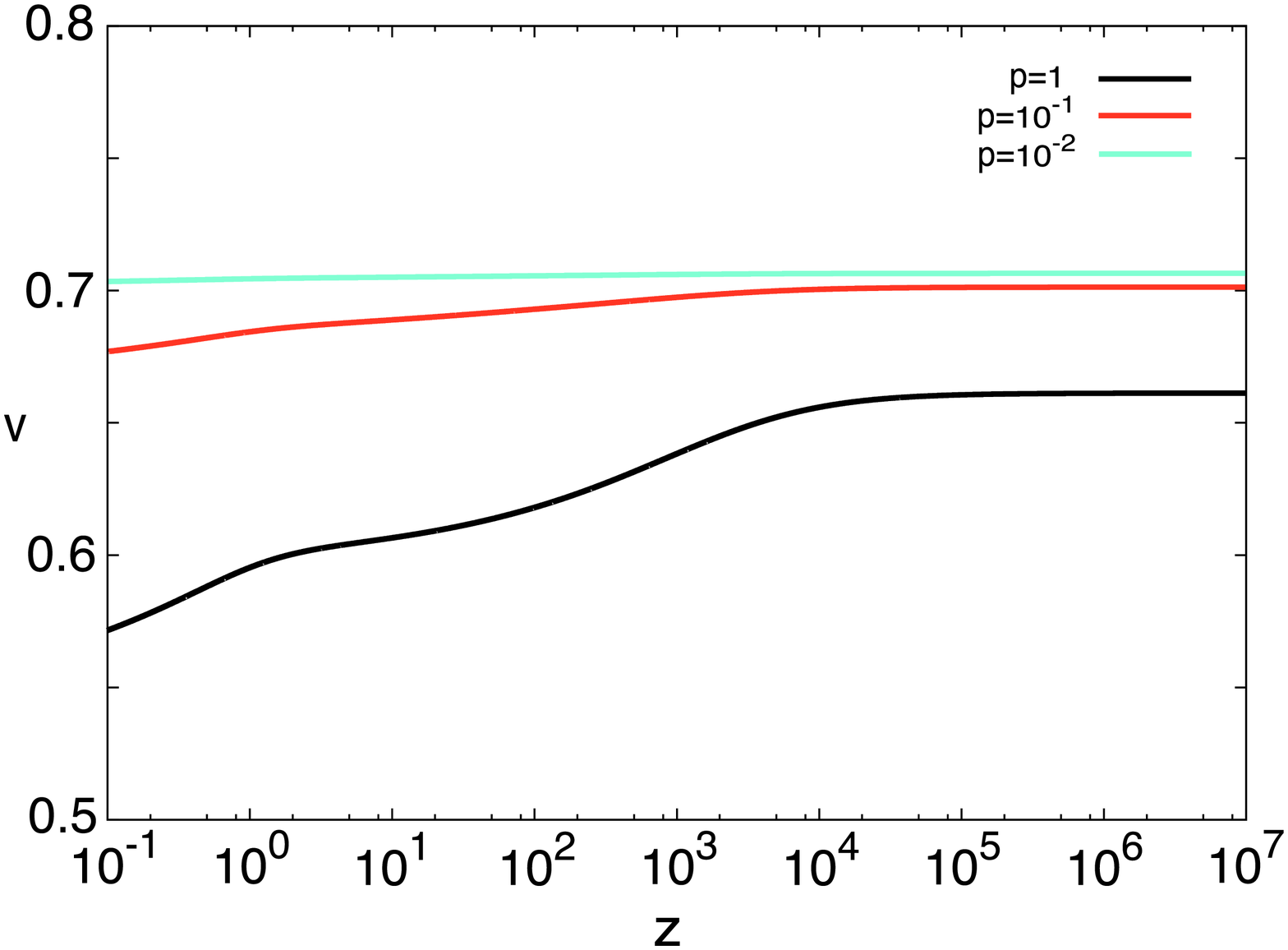}
\label{fig:v}
}

\end{center}
\end{minipage}
\caption{The time evolution of $\gamma$ and $v$ for 
  different values of $p$.}
\end{figure}

Figs. \ref{fig:gamma} and \ref{fig:v} show the time evolutions of
$\gamma(t)$ and $v(t)$ and their dependence on the reconnection
probability $p$.  For $p=1$, $\gamma$ and $v$ reach the asymptotic
values in the matter-dominated era from those in the
radiation-dominated era by $z\sim 10^2$.  After the dark energy
becomes dominant component of the universe at $z\lesssim 1$, $\gamma$
begins to increase and $v$ begins to decrease, because the exponential
expansion of the universe dilutes strings and makes them slow down.
For small $p$, we see the overall magnitude of $\gamma$ decreases as
$p$ decreases, the dependence is proportional to $p^{-1}$ and
$p^{-1/2}$ for the radiation- and matter-dominated era, respectively
\cite{Kuroyanagi:2012wm}.  The asymptotic value of $v$ is
approximately $1/\sqrt{2}$ in both the radiation- and matter-dominated
eras.

\subsection{Gravitational waves from cosmic string loops}

Here, we briefly describe the formalism to calculate GWs from cosmic
strings.  The detail is described in our previous paper
\cite{Kuroyanagi:2012wm}, or originally in Refs.
\cite{Damour:2000wa,Damour:2001bk,Damour:2004kw,Siemens:2006yp}.

Since the main source of GWs in the string network is cusps on loops,
we first evaluate the density of loops in the Universe in order to
calculate GWs from them.  In the scaling regime, where the typical
length scale of infinite strings is proportional to the Hubble scale
and their number in a Hubble horizon remains constant, infinite
strings continuously convert their length into loops.  The number
density of loops formed at time $t_i$ is given by
\beq
\frac{dn}{dt_i} (t,t_i)dt_i=\frac{dt_i}{\alpha \gamma(t_i)^2 t_i^4}\left (\frac{a (t_i)}{a (t)}\right)^3,
\label{loopdensity}
\eeq
at time $t$.  Loops continue to shrink by releasing energy as GWs and
eventually evaporate.  The length of a loop formed at $t_i$ is given
by
\beq
l (t,t_i)=\alpha t_i - \Gamma G\mu  (t-t_i), \label{length}
\eeq
for time $t$.  Here, $\Gamma$ is a constant which represents the
efficiency of the GW emission from loops and set to $50$ in this
paper.  If $\alpha \gg \Gamma G\mu$, loops are long-lived, that is,
they survive more than a Hubble time.  On the other hand, if $\alpha
\ll \Gamma G\mu$, loops are short-lived, and they evaporate within a
Hubble time.  Using Eqs. (\ref{loopdensity}) and (\ref{length}), we
can describe the number density of loops in terms of length $l$ and
time $t$.

Cusps emit GWs of $f\gg l^{-1}$ into a small solid angle.  The
linearly polarized waveform of a GW burst emitted in a direction
$\mathbf{n}$ by a loop with length $l$ at redshift $z$ is given by
\beq
h_{\mu\nu} (t,\mathbf{n})=\int df h (f,z,l) e^{-2\pi i ft}e^+_{\mu\nu} (\mathbf{n}) \times \Theta (\mathbf{n}\cdot \mathbf{n}_c - \cos\left[\theta_m (f,z,l)\right]) \times \Theta (1-\theta_m (f,z,l)), \label{waveform}
\eeq
where $\mathbf{n}_c$ is the direction of the center of the burst,
$\theta_m$ is the beaming angle of the GW burst which is given by
\beq
\theta_m (f,z,l)= ( (1+z)fl)^{-1/3}, \label{thetam}
\eeq
and $e^+_{\mu\nu}=l_{\mu}m_{\nu}-l_{\nu}m_{\mu}$ is the polarization
tensor (for plus polarization), where $l_{\mu}= (0,\mathbf{l}),
m_{\mu}= (0,\mathbf{m})$ and $\mathbf{l}$ and $\mathbf{m}$ are unit
vectors orthogonal to $\mathbf{n}$ and each other.  The Fourier
transform of the GW amplitude, $h(f,z,l)$, is given by
 \beq
h (f,z,l)\approx 2.68\frac{G\mu l}{ ( (1+z)fl)^{1/3}r (z)f}, \label{strain}
\eeq
where $r (z)=\int^z_0 dz^{\prime}/H (z^{\prime})$.  The first Heviside
step function $\Theta$ in Eq. (\ref{waveform}) is introduced to
account for the beaming of the GW and the second one is for the low
frequency cutoff at $f\lesssim l^{-1}$.

The arrival rate of GW bursts with frequency $f$ and amplitude $h$
emitted at redshift $z$ is given by
\beq
\frac{dR}{dzdh} (f,h,z)=\frac{3}{4}\theta^2_m (f,z,l)\frac{c}{ (1+z)h}\frac{1}{\gamma(t_i)^2 \alpha t^4_i}\frac{1}{\alpha+\Gamma G\mu}\left (\frac{a (t_i)}{a (t)}\right)^3 \frac{dV}{dz}\Theta  (1-\theta_m (f,z,l)),
\label{dR_dzdh}
\eeq
where
\beq
\frac{dV}{dz} (z)=\frac{4\pi a^2 (z)r^2 (z)}{H (z) (1+z)}.
\eeq
From Eqs. (\ref{length}) and (\ref{strain}), $l$ and $t_i$ can be
expressed as
\beq
l (f,h,z)=\left (\frac{hr (z)}{2.68G\mu} (1+z)^{1/3}f^{4/3}\right)^{3/2}, \label{l_of_hzf}
\eeq
\beq
t_i (f,h,z)=\frac{l (f,h,z)+\Gamma G\mu t (z)}{\alpha+\Gamma G\mu}, \label{t_i}
\eeq
which enables to express Eq. (\ref{dR_dzdh}) in terms of $h$, $z$, and
$f$.  Finally, the total arrival rate of GWs today for given frequency
and amplitude is given by
\beq
\frac{dR}{dh}=\int^{\infty}_0 dz \frac{dR}{dhdz}.
\eeq

The GW bursts are identified as a single burst if they do not overlap
each other.  In contrast, bursts overlapping each other are observed
as a GW background.  Namely, a GW background is formed by bursts which
come to the observer with a time interval shorter than the oscillation
period of themselves.  According to the criteria in
Ref. \cite{Siemens:2006yp}, such bursts have amplitude smaller than
$h_*$, which is determined for a given frequency as
\beq
\int^{\infty}_{h_*} dh\frac{dR}{dh}=f.
\eeq
Then, the amplitude of a GW background, $\Omega_{\rm GW} (f)\equiv
(d\rho_{\rm GW}/d\ln f)/\rho_{cr}$, where $\rho_{\rm GW}$ is the
energy density of the GWs and $\rho_{cr}$ is the critical density of
the universe, is given by
\beq
\Omega_{\rm GW} (f)=\frac{2\pi^2}{3H_0^2}f^3\int^{h_*}_0dhh^2\frac{dR}{dh}.
\label{OmegaGW}
\eeq
The main contribution to $\Omega_{\rm GW}$ at frequency $f$ comes from
loops expiring at the redshift which satisfies $\min\{\alpha,\Gamma
G\mu\}\times t\sim f^{-1}(1+z)^{-1}$ in the radiation-dominated era or
loops expiring recently.

Here, we briefly mention the parametric dependence of the burst rate
and $\Omega_{\rm GW}$.  The burst rate and the amplitude of the GW
background are basically enhanced as $G\mu$ increases.  However, the
value of $G\mu$ also affects the spectral shape of the burst rate and
GW background, since the number density of loops are affected by
$G\mu$ through the lifetime of loops.  This is important especially in
the case of $G\mu>\alpha$, where $G\mu$ determines the size of
expiring loops.  The large value of $\alpha$ decreases the initial
number density of loops.  However, it also has an effect to increase
the number density of loops, since large $\alpha$ makes the lifetime
of loops longer.  Therefore, the effects of $\alpha$ on the burst rate
and the background depend on values of $f$, $h$ and the other
parameters.  Larger $p$ simply leads to a larger burst rate and larger
amplitude of the background, through the factor $\gamma^{-2}$ in Eq.
(\ref{dR_dzdh}).  We refer to Ref. \cite{Kuroyanagi:2012wm} for more
detailed discussion.

\subsection{CMB fluctuations induced by cosmic strings}

Here, we describe the method to calculate the power spectra of CMB
fluctuations produced by cosmic strings.  We refer to Refs.
\cite{Pogosian:1999np, CMBACT} for the details.

The evolution of the string network is highly non-linear and produce
all types of perturbations: scalar, vector and tensor modes.  For the
calculation of the CMB power spectra, we use a code based on CMBACT
\cite{CMBACT}, which we modify to include the time evolution of
$\gamma$ and $v$.  This code is based on the semi-analytical method
described in Ref. \cite{Pogosian:1999np}, which models the string
network as an ensemble of discrete straight line segments.  The length
and the velocity of each segment are set to the solution of
Eqs. (\ref{gammaeq}) and (\ref{veq}).  The position and the direction
of the velocity are randomly selected for each segment.  At each time
step, some segments are removed so that the number density of strings
is consistent with the scaling.  Then, we can derive the
energy-momentum tensor of such a simplified network and compute the
power spectra of CMB fluctuations.  In this paper, we assume that
infinite strings have no wiggliness, which is introduced in
Ref. \cite{Pogosian:1999np}).

\begin{figure}[t]
\begin{tabular}{cc}
\begin{minipage}{0.5\hsize}
\begin{center}
  \subfigure[Dependence of the temperature spectrum on $G\mu$.  The
  green, black and red solid lines represent the spectra for
  $G\mu=10^{-6.5}$, $G\mu=10^{-7}$ and $G\mu=10^{-7.5}$, respectively,
  with $p$ fixed to 1.  ]{
\includegraphics[width=75mm]
{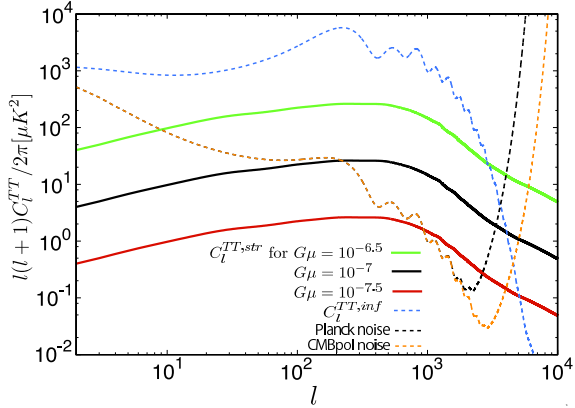}
\label{fig:CT_Gmu}
}
\end{center}
\end{minipage}

\begin{minipage}{0.5\hsize}
\begin{center}
  \subfigure[Dependence of the temperature spectrum on $p$.  The
  black, red and green solid lines represent the spectra for $p=1$,
  $p=10^{-0.5}$ and $p=10^{-1}$, respectively, with $G\mu$ fixed to
  $10^{-7}$.  ]{
\includegraphics[width=75mm]
{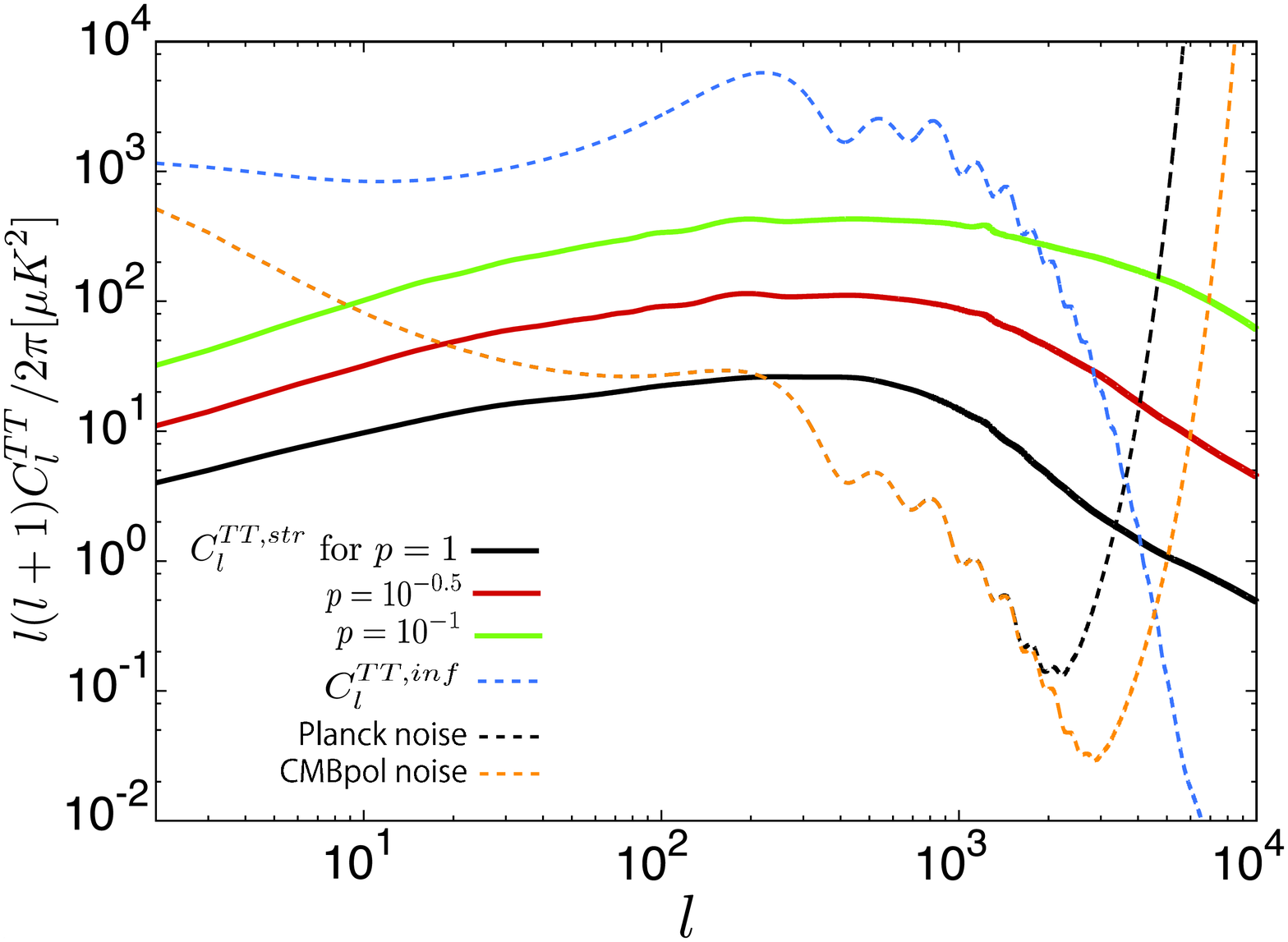}
\label{fig:CT_p}
}

\end{center}
\end{minipage}
\end{tabular}
\caption{The power spectra of CMB temperature fluctuations induced by
  strings for various parameter sets.  In each figure, the black solid
  line represents the spectrum for the fiducial parameter set taken
  for the Fisher analysis in Sec.  4.3.  We also show the spectrum
  predicted by inflation as a blue dotted line and the expected noise
  levels for Planck and CMBpol as a black and orange dotted line,
  respectively.}
\label{fig:CT}
\end{figure}

\begin{figure}[t]
\begin{tabular}{cc}
\begin{minipage}{0.5\hsize}
\begin{center}
  \subfigure[Dependence of the B-mode spectrum on $G\mu$.  The green,
  black and red solid lines represent the spectra for
  $G\mu=10^{-6.5}$, $G\mu=10^{-7}$ and $G\mu=10^{-7.5}$, respectively,
  with $p$ fixed to 1.  ]{
\includegraphics[width=75mm]
{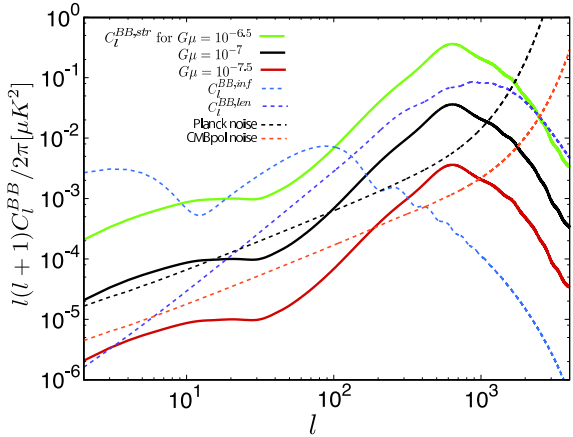}
\label{fig:CB_Gmu}
}
\end{center}
\end{minipage}

\begin{minipage}{0.5\hsize}
\begin{center}
  \subfigure[Dependence of the B-mode spectrum on $p$.  The
  black, red and green solid lines represent the spectra for $p=1$,
  $p=10^{-0.5}$ and $p=10^{-1}$, respectively, with $G\mu$ fixed to
  $10^{-7}$.  ]{
\includegraphics[width=75mm]
{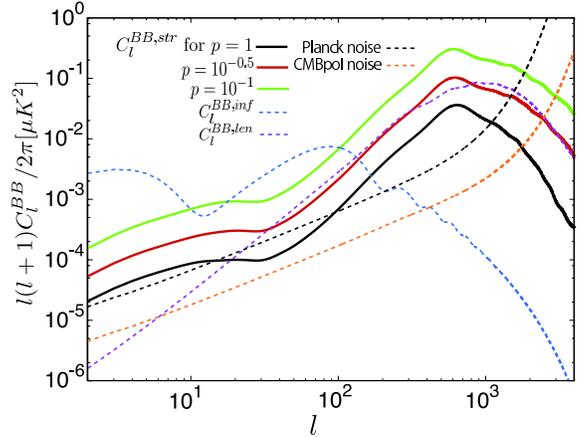}
\label{fig:CB_p}
}
\end{center}
\end{minipage}
\end{tabular}
\caption{The power spectra of CMB B-mode polarization induced by
  strings for various parameter sets.  In each figure, the black solid
  line represents the spectrum for the fiducial parameter set taken in
  the Fisher analysis in Sec.  4.3.  The blue dotted line is the
  spectrum induced by the inflationary GWs.  The tensor-to-scalar
  ratio $r$ is set to $0.1$.  The purple dotted line is the spectrum
  of B-mode polarization generated through the gravitational lensing
  of the inflationary E-mode polarization.  The black and orange
  dotted line represent the expected noise levels for Planck and
  CMBpol, respectively.  }
\label{fig:CB}
\end{figure}

Here, we briefly discuss the parametric dependencies of the CMB power
spectra.  More detailed discussion is given in
Ref. \cite{Pogosian:2007gi}.  In Figs. \ref{fig:CT} and \ref{fig:CB},
we show the CMB power spectra of temperature fluctuations
$C^{TT,str}_l$ and B-mode polarization $C^{BB,str}_l$ induced by
cosmic strings, respectively, for various parameter sets.  We also
show the temperature spectrum induced by the inflationary primordial
perturbations $C^{TT,inf}_l$ in Fig. \ref{fig:CT}, the B-mode
polarization from the inflationary GWs $C^{BB,inf}_l$ and that from
the gravitational lensing of E-modes $C^{BB,len}_l$ in
Fig. \ref{fig:CB}, which are calculated by
CAMB\cite{Lewis:1999bs,CAMB}.  In addition, we plot the expected noise
levels of Planck and CMBpol, which is given by the sum of the
instrumental noise and the cosmic variance
\beq
  \frac{l(l+1)}{2\pi}N^{\rm
    tot}_{T;l}=\frac{l(l+1)}{2\pi}\sqrt{\frac{2}{(2l+1)l}}(C^{TT,inf}_l+N_{T;l}), \label{totnoiseT}
\eeq 
for temperature fluctuations, and 
\beq 
\frac{l(l+1)}{2\pi}N^{\rm
    tot}_{B;l}=\frac{l(l+1)}{2\pi}\sqrt{\frac{2}{(2l+1)l}}(C^{BB,len}_l+N_{P;l}),
\label{totnoiseB}
\eeq 
for B-modes.  \footnote{The variance of $C^i_l$ is given by $(\Delta
  C^i_l)^2=\frac{2}{2l+1}(C^{i}_l+N_{a;l})^2$, where $i$ denotes $TT$
  or $BB$ and $a$ denotes $T$ or $P$.  When we take a logarithmically
  homogeneous binning of $l$ with bin width $\Delta\ln l=1$, there are
  $l$ multipoles in a bin at $l$.  Since different multipoles are
  independent, the noise level par each bin should be given by $\Delta
  C^i_l/\sqrt{l}$.  This allows us rough estimation of the
  detectability of the signal by eye in Figs. \ref{fig:CT} and
  \ref{fig:CB}. } The instrumental noise spectra for the temperature
$N_{T;l}$ and the polarization $N_{P;l}$ are defined in Sec. 3.

The power spectrum of CMB temperature fluctuations induced by strings
has a single bump, unlike the acoustic oscillations seen in the
inflationary CMB power spectra.  This is because strings generate
fluctuations constantly and such fluctuations are incoherent, while
the primordial fluctuations generated during inflation oscillate
coherently.  The peak of $C^{TT,str}_l$ corresponds to the scale of
the perturbations generated at last scattering.  Another important
difference is that $C^{TT,str}_l$ does not decay exponentially as $l$
increases in contrast to the Silk dumping of the inflationary spectrum
at large $l$.  This makes $C^{TT,str}_l$ larger than $C^{TT,inf}_l$ at
high $l$ and it may be observable by future experiments\footnote{ The
  power spectra of E-mode polarization $C^{EE}_l$ and the
  cross-correlation between temperature and E-mode $C^{TE}_l$ by
  strings also decay more slowly than the inflationary spectrum at
  high $l$.  However, they have small amplitude and do not affect the
  observation when we assume realistic parameters.}.  The B-mode power
spectrum also does not have oscillations, but has two bumps.  The bump
at high $l$ and that at low $l$ correspond to the polarization
generated around last scattering and around reionization,
respectively.  If the amplitude of the primordial gravitational wave
is small, strings can be the main source of B-mode polarization.

The dependence of CMB power spectra on $G\mu$ is simple.  The
amplitudes of temperature and polarization fluctuations are
proportional to $G\mu$, so the power spectra are proportional to
$(G\mu)^2$.  The effect of $p$ arises through the correlation length
of the string network $\gamma$ and the r.m.s. velocity of strings $v$.
For small $p$, $\gamma$ becomes smaller and $v$ becomes larger.  If
$p$ is small, the small value of $\gamma$ makes the string network
denser and enhances the amplitude of the CMB fluctuations.  Also, the
small correlation length $\gamma$ makes the typical scale of
perturbations smaller.  However, at the same time, large $v$ makes the
typical scale of perturbations larger \cite{Pogosian:2007gi}.  Since
the two effects on the typical scale of perturbations compensate, we
find no apparent shift of the peak in Figs. \ref{fig:CT} and
\ref{fig:CB} with the change of the value of $p$.  While these effects
tend to compensate, small effects are seen in the width of the peak
and the slope of the small-scale spectrum.

Since the CMB anisotropy is induced not by loops but by infinite strings,
the CMB power spectra do not depend on $\alpha$.

\section{Formalism to evaluate the sensitivity and the constraints on
  parameters in each experiment}

In this section, we describe the method to calculate the Fisher
information matrix, which is used to estimate the power of each
experiment to determine the string parameters.  For GW burst and
background detection by interferometers, Ref. \cite{Kuroyanagi:2012wm}
provides more detailed derivations.

Given a data-set of experiments, the parameter values that are most
likely to result in the model prediction are those which maximize the
likelihood function ${\cal L}$.  Such a method for parameter
estimation is called the maximum likelihood method and widely used in
the analysis of cosmological observations \cite{Tegmark:1996bz}.  The
error in this estimation can be predicted by calculating the Fisher
information matrix
\begin{equation}
{\cal F}_{lm}\equiv -\frac{\partial^2\ln{\cal L}}{\partial\theta_l\partial\theta_m}.
\label{Fisher}
\end{equation}
Assuming a Gaussian likelihood, the expected error in the parameter
$\theta_l$ is given by 
\begin{equation}
\sigma_{\theta_l}=\sqrt{ ({\cal F}^{-1})_{ll}}.
\end{equation}

\subsection{GW burst detection}
The search for GW bursts signals from cosmic strings is performed by
matched filtering \cite{Siemens:2006vk,Abbott:2009rr}, where we assume
that the burst signal from a cosmic string cusp is linearly polarized
and have the frequency dependence of $f^{-4/3}$.  Then, the spectrum
form is expressed by
\begin{equation}
h^+ (f)=Af^{-4/3}\Theta (f_h-f)\Theta (f-f_l),
\end{equation}
where the amplitude $A$ can be read from Eq. (\ref{strain}).  The low
frequency cutoff $f_l$ is given by the low frequency limit of the
experiment and the high frequency cutoff $f_h$ is typically given by
the most sensitive frequency of the detector \cite{Cohen:2010xd}.  The
signal to noise ratio(SNR) $\rho$ is given by
\begin{equation}
\rho=\left[4\int^{f_h}_{f_l}df\frac{|\hat h (f)|^2}{S_n (f)}\right]^{1/2},
\end{equation}
where the GW signal $\hat h(f)=F^+h^+(f)$ is given by multiplying the
detector response to plus polarized GWs $F^+$.  Effectively, $F^+$ can
be replaced with the all sky-averaged value for orthogonal arm
detectors, $\overline{F^+}\sim 1/\sqrt{5}$ for a single detector, or
$\overline{F^+}\sim 1$ for the GW detector network which has $100\%$
visibility over the whole sky.  The noise spectral density $S_n (f)$
is defined by $\langle n (f)^*n (f)\rangle\equiv S_n (f)\delta
(f-f^\prime)/2$, where $n(f)$ is the Fourier transform of the detector
noise $n(t)$ and $\langle\cdots\rangle$ denotes the ensemble average.
In this paper, we take the detection threshold as $\rho>4$
\cite{Damour:2000wa}.

For current LIGO, the noise spectrum is given by 
\begin{eqnarray}
S_n (f)=
1.09\times 10^{-41}\left (\frac{30{\rm Hz}}{f}\right)^{28}+1.44\times 10^{-45}\left (\frac{100{\rm Hz}}{f}\right)^4\nonumber\\
+1.28\times 10^{-46}\left (1+\left (\frac{90{\rm Hz}}{f}\right)^{-2} \right)
{\rm Hz}^{-1}.
\label{LIGO}
\end{eqnarray}
and we take $f_l=40$Hz and $f_h=150$Hz \cite{Siemens:2006vk}.  The
detection threshold $\rho>4$ corresponds to the detection limit
$A\simeq 9.1\times 10^{-21}{\rm s}^{-1/3}$ or $fh\simeq 1.7\times
10^{-21}$ at $f=150{\rm Hz}$.  For the future ground-based GW
detectors, such as Advanced-LIGO, we use
\begin{equation}
S_n (f)=10^{-49}\left[x^{-4.14}-\frac{5}{x^2}+111\left (\frac{2-2x^2+x^4}{2+x^2}\right)\right]{\rm Hz}^{-1},
\label{Adv-LIGO}
\end{equation}
where $x=f/ (215{\rm Hz})$ \cite{Key:2008tt}, and we take $f_l=10$Hz
and $f_h=220$Hz.  If we consider a world-wide detector network and
assume $\overline{F^+}\sim 1$, the detection limit is $A\simeq
2.1\times 10^{-22}{\rm s}^{-1/3}$ or $fh\simeq 3.4\times 10^{-23}$ at
$f=220{\rm Hz}$.  For eLISA/NGO, we use
\begin{equation}
S_n (f)=\frac{20}{3}\frac{4S_{\rm acc}+S_{\rm sn}+S_{\rm omn}}{L^2}\left(1+\left(\frac{f}{0.41(\frac{c}{2L})}\right)^2\right),
\label{eLISA}
\end{equation}
where $S_{\rm acc}=1.37\times 10^{-32}\left(1+\frac{10^{-4}{\rm
      Hz}}{f}\right)f^{-4}{\rm m}^2{\rm s}^{-4}{\rm Hz}^{-1}$, $S_{\rm
  sn}=5.25\times 10^{-23}{\rm m}^2{\rm Hz}^{-1}$ and $S_{\rm
  omn}=6.28\times 10^{-23}{\rm m}^2{\rm Hz}^{-1}$ with the arm length
$L=1.0\times 10^6$km \cite{AmaroSeoane:2012km}, a take $f_l=10^{-6}$Hz
and $f_h=7.0\times 10^{-3}$Hz.  Then, the detection limit is $A\simeq
8.9\times 10^{-22}{\rm s}^{-1/3}$ or $fh\simeq 5.8\times 10^{-21}$ at
$f=7.0\times10^{-3}{\rm Hz}$.  For BBO or DECIGO, whose sensitivities
are roughly the same, we use the configuration of BBO.  It is designed
to use a technique called time-delay interferometry, and the noise
spectrum is given by
\begin{equation}
S_n (f)=\left(\frac{R_A(f)}{S_A(f)}+\frac{R_E(f)}{S_E(f)}+\frac{R_T(f)}{S_T(f)}\right)^{-1}.
\label{BBODECIGO}
\end{equation}
The subscripts (A,E,T) denote the TDI variables and each noise spectrum
is given by
\begin{eqnarray}
     &&S_{A}(f)=S_{E}(f)=8\sin^2(\hat{f}/2)[(2+\cos\hat{f})S_{\rm shot}
     +2(3+2\cos\hat{f}+\cos(2\hat{f}))S_{\rm accel}],\nonumber\\
     &&S_T(f)=2[1+2\cos\hat{f}]^2[S_{\rm shot}+4\sin^2(\hat{f}/2)S_{\rm accel}],
\end{eqnarray}
where $S_{\rm shot}=2.0\times 10^{-40}/(L/{\rm km})^{-2}{\rm Hz}^{-1}$
and $S_{\rm accel}=9.0\times 10^{-40}/(2\pi f/{\rm Hz})^{-4}/(2L/{\rm
  km})^{-2}{\rm Hz}^{-1}$ with the arm length $L=5.0\times 10^4$km.
For the calculation of the detector response $R_{A,E,T}$, see
Ref. \cite{Cornish:2002rt}.  We take $f_l=0.1$Hz and $f_h=0.25$Hz,
where the low-frequency cutoff is determined to take into account the
confusion noise from white dwarf binaries.  This leads to the
detection limit $A\simeq 1.2\times 10^{-24}{\rm s}^{-1/3}$ or
$fh\simeq 1.9\times 10^{-24}$ at $f=0.25{\rm Hz}$.

Let us suppose that we detect a sufficient number of GW bursts at
$f_{\rm best}$, at which the detector is most sensitive.  Under the
assumption that the number of bursts follows a Poisson
distribution \cite{gr-qc/0312056,arXiv:0710.0497,arXiv:1004.3499}, the
Fisher matrix is given by
\begin{equation}
{\cal F}_{lm}=\int^{\infty}_{h_{\rm min}}\frac{\partial\Phi}{\partial\theta_l}\frac{\partial\Phi}{\partial\theta_m}\frac{1}{\Phi}dh,
\end{equation}
where $\Phi(h)\equiv dR/dh\times T$, which is a function of the model
parameters, $T$ is the observation time and $h_{\rm min}$ is the
smallest amplitude of the detectable burst at $f=f_{\rm best}$.

\subsection{Search for the stochastic GW background by interferometers}

The GW background is searched by correlating output signals of two or
multiple interferometers.  The SNR in such a correlation analysis with
$N$ detectors is given by \cite{Allen:1997ad,Kudoh:2005as}
\begin{equation}
\rho=
\left[\sum^N_{I=1}\sum^N_{J<I}\rho_{IJ}^2\right]^{1/2},
\end{equation}
where
\beq
\rho_{IJ}=\frac{2}{5}\sqrt{2T}\left[ \int^\infty_{0}df\frac{S_h(f)^2|\gamma_{IJ}(f)|^2}{R_{IJ}(f)}\right]^{1/2}, \label{rhoIJexact}
\eeq
\beq
R_{IJ}(f)=\left(\frac{2}{5}\right)^2S_h(f)^2\left(|\gamma_{IJ}(f)|^2+\gamma_{II}\gamma_{JJ}\right)+\frac{2}{5}S_h(f)\left(\gamma_{II}(f)S_{n,J}(f)+\gamma_{JJ}S_{n,I}(f)\right)+S_{n,I}(f)S_{n,J}(f),
\eeq
and $T$ is the observation time.  The subscripts $I$ and $J$ refer to
independent signals obtained at each detector or TDI variables
(A,E,T).  The overlap reduction function between the $I$-th and $J$-th
detector $\gamma_{IJ}$ is given by
\begin{equation}
\gamma_{IJ} (f)\equiv\frac{5}{8\pi}\int d\hat{\bf \Omega}
 (F^+_I(f,\hat\Omega)F^+_J(f,\hat\Omega)+F^{\times}_I(f,\hat\Omega)F^{\times}_J(f,\hat\Omega))
e^{-2\pi if\hat{\bf \Omega}\cdot ({\bf x}_I-{\bf x}_J)},
\end{equation}
where $F^+_I$ or $F^{\times}_I$ is the detector response to plus or
cross polarized GWs of the $I$-th detector, $\mathbf{x}_I$ is the
position of the $I$-th detector and $\hat\Omega$ is the direction of
GWs.  We calculate $\gamma_{IJ}$ following the procedure 
given in Ref.  \cite{Nishizawa:2009bf} for ground-based detectors and
Ref. \cite{Corbin:2005ny} for BBO. 
The signal spectrum can be corresponded to $\Omega_{\rm GW}$ as
\beq
S_h(f)=\frac{3H^2_0}{4\pi^2f^3}\Omega_{\rm GW}(f).
\eeq
In the weak signal approximation, $S_h(f)\ll S_{n,I}(f)$,
Eq. (\ref{rhoIJexact}) reduces to
\begin{equation}
\rho_{IJ}\simeq\frac{3H_0^2}{10\pi^2} \sqrt{2T}
\left[\int^{\infty}_{0}df\frac{|\gamma_{IJ} (f)|^2
\Omega_{\rm GW} (f)^2}{f^6S_{n,I} (f)S_{n,J} (f)}\right]^{1/2}.
\label{rhoIJ}
\end{equation}
The weak signal approximation is valid for $\rho \lesssim
200$ \cite{Kudoh:2005as,Seto:2005qy}.

From the non-detection of GW background, we can put an upper limit on
$\Omega_{\rm GW}$.  Current LIGO detectors set an upper bound of
$\Omega_{\rm GW}<7.2\times10^{-6}$ for $41.5{\rm Hz}<f<169.25{\rm
  Hz}$, assuming a flat spectrum.  A 3-year run of the future detector
network including Advanced LIGO, Advanced Virgo and KAGRA, would reach
$\Omega_{\rm GW}= 4.5\times10^{-9}$ at $10{\rm Hz}<f<200{\rm Hz}$.  A
3-year run of BBO/DECIGO would provide $\Omega_{\rm GW}=
9.2\times10^{-17}$ for $0.1{\rm Hz}<f<10{\rm Hz}$.  Since eLISA/NGO is
designed to have only one independent channel, we do not consider the
cross correlation analysis in eLISA/NGO.

The Fisher matrix for the GW background measurement is generally given by
\beq
\mathcal{F}_{lm}=\sum^N_{I=1}\sum^N_{J<I}\frac{8T}{25}\int^{\infty}_0df\frac{|\gamma_{IJ}(f)|^2\partial_{\theta_l}S_h(f)\partial_{\theta_m}S_h(f)}{R_{IJ}(f)}.
\label{FGWBexact}
\eeq
Under the weak signal approximation, it reduces to
\cite{Abbott:2003hr,Seto:2005qy}
\begin{equation}
{\cal F}_{lm}=\left (\frac{3H_0^2}{10\pi^2}\right)^2 2T\sum^N_{I=1}\sum^N_{J<I}
\int^{\infty}_{0}df\frac{|\gamma_{IJ} (f)|^2\partial_{\theta_l}
\Omega_{\rm GW} (f)\partial_{\theta_m}\Omega_{\rm
GW} (f)}{f^6S_{n,I} (f)S_{n,J} (f)}.
\label{FGWBweak}
\end{equation}

\subsection{Search for the stochastic GW background in pulsar timing
  experiments}
Pulsar timing experiments provide a unique opportunity to observe GWs
in low-frequency band $10^{-9}-10^{-7}{\rm Hz}$
\cite{Estabrook,Sazhin,Detweiler} (for a review, see
Ref. \cite{Lorimer}).  The analysis is based on the measurement of
pulse time-of-arrival (TOA) variations.  The stochastic GW background
causes fluctuations in the TOAs.  One can extract the signal from
noise associated with individual pulsars by correlating TOAs between
different pulsars.

We follow the formalism described in Ref.  \cite{Jenet:2005pv} to
calculate the SNR and the Fisher matrix for detection of GWs in pulsar
timing experiments.  Let us assume observations of $M\gg 1$ pulsars at
time $t_0,t_1,...,t_{N-1}$ with the time interval $\Delta t$.  The
total observation time is $T=N\Delta t$ and $N\gg 1$.  Then, we can
make $N_p=M(M-1)/2$ pulsar pairs from $M$ pulsars. We denote the
timing residual of $i$-th pulsar at time $t_a$ as $R_i(t_a)$.  The
correlation coefficient of $i$-th pair is defined as \beq r_i\equiv
\frac{1}{N}\sum^{N-1}_{a=0}R_{i_1}(t_a)R_{i_2}(t_a), \eeq where $i_1$
and $i_2$ is the number allocated to the first and second pulsar in
the $i$-th pair.  Under the existence of the isotropic stochastic GW
background, the ensemble average of $r_i$ is\cite{Hellings} \beq
\left<r_i\right>=\sigma_g^2\zeta(\theta_i), \eeq where $\sigma_g$ is
the root mean square of the timing residuals induced by the GW
background and given by 
\beq 
\sigma_g^2=\int^{f_h}_{f_l}P_R(f)df.
\eeq
The highest and lowest frequency of GWs are given by $f_h=1/2\Delta t$
and $f_l=1/T$.  The power spectrum of the timing residuals $P_R$ is
defined as $\left<
  R_i(f)R_i(f^{\prime})\right>=P_R(f)\delta(f-f^{\prime})/2$ and that
induced by the GW background is given by 
\beq
P_R(f)=\frac{H_0^2}{8\pi^4f^5}\Omega_{\rm GW}(f).  
\eeq 
Note that $r_i$ induced by the GW background has a specific dependence
on the angle between the direction to $i_1$-th and $i_2$-th pulsars,
$\theta_i$.  This dependence is characterised by $\zeta(\theta_i)$,
which is described as 
\beq
\zeta(\theta)=\frac{3}{2}x\ln x-\frac{x}{4}+\frac{1}{2}(1+\delta(x)),
\ x=\frac{1-\cos\theta}{2}, \ \delta(x)=
\begin{cases}
& 1 \ ; \ {\rm for}  \ x=0 \\
& 0 \ ; \ {\rm otherwise} 
\end{cases}.
\label{zeta}
\eeq 
The signature of the GW background can be extracted by calculating the
following quantity, 
\beq
S=\frac{\frac{1}{N_p}\sum^{N_p-1}_{i=0}(r_i-\bar
  r)(\zeta(\theta_i)-\bar\zeta)}{\sigma_r\sigma_\zeta},
\label{Spul}
\eeq 
where $\bar r$ and $\bar \zeta$ are the arithmetic mean over all pairs
of pulsars, and $\sigma_r^2$ and $\sigma_\zeta^2$ are the sample
variance of $r$ and $\zeta$.  We define this quantity as a ``signal".
If pulsars are distributed isotropically, $\bar\zeta=0$ and
$\sigma_\zeta^2=1/\sqrt{48}$.  Hereafter, we replace $\bar\zeta$ and
$\sigma_\zeta^2$ by these values.  The ensemble average of $S$ is
given by
\beq
\left<S\right>=\frac{\sigma^2_g\sigma_\zeta}{\sqrt{\sigma^4_g\sigma^2_\zeta+\sigma^2_{\Delta
      r}}}, \eeq where \beq \sigma^2_{\Delta
  r}=\frac{1}{N_p}\sum^{N_p-1}_{i=0}\left<(r_i-\left<r_i\right>)^2\right>.
\eeq
If there is no correlation between data of different pulsars,
$\left<r_i\right>=0$, $S$ follows a Gaussian distribution with zero
mean and variance $1/N_p$.  Therefore, we define the SNR as
$\rho\equiv\left<S\right>\sqrt{N_p}$.

We assume that the noise of each pulsar is white and uncorrelated with
that of the other pulsar and the signal of GWs.  We also assume that
all pulsars have the same noise and denote the root mean square of
time residuals induced by noise as $\sigma_n$.  Then, the SNR becomes
\beq
\rho=\sqrt{\frac{M(M-1)/2}{1+\left[\chi(1+\bar{\zeta^2})+2(\sigma_n/\sigma_g)^2+(\sigma_n/\sigma_g)^4\right]/N\sigma^2_{\zeta}}},
\label{rhopul}
\eeq
where
\beq
\chi=\frac{1}{N\sigma_g^4}\sum_{a=0}^{N-1}\sum_{b=0}^{N-1}c_{ab}^2, \ c_{ab}=\left<R_i(t_a)R_i(t_b)\right>.
\eeq

Assuming that $P_R(f)$ is the monotonically decreasing function of
$f$, the SNR can be enhanced by low-pass filtering and whitening,
which modifies the SNR to \cite{Jenet:2005pv}
\beq
\tilde\rho=\sqrt{\frac{M(M-1)/2}{1+\tilde{\sigma}^2_{\Delta r}/\tilde{\sigma}_g^4\sigma_\zeta^2}},
\label{rhopul2}
\eeq
where
\beq
\tilde{\sigma}_g^2=\frac{2}{N}\sigma^2_d\sum^{N_{\rm max}}_{i=1}\frac{P_R(i)}{P_d(i)}, \ \tilde{\sigma}_{\Delta r}^2=\frac{2\sigma^4_d}{N^2}\sum^{N_{\rm max}}_{i=1}
\left(1+\left(\frac{P_R(i)}{P_d(i)}\right)^2\bar{\zeta^2}\right).
\eeq
Here, we define the discrete power spectra for the $i$-th frequency
bin as
\beq
 P_g(i)\equiv\int^{f_{i+1}}_{f_i}dfP_R(f), \ P_n(i)\equiv \int^{f_{i+1}}_{f_i}dfP_n(f)=\frac{2\sigma_n}{N}, \ P_d(i)\equiv P_g(i)+P_n(i),
\eeq
where
\beq
  f_i=
  \begin{cases}
  \frac{0.97}{T} \ ; \ {\rm for} \ i=1 \\
  \frac{i-0.5}{T}\ ; \ {\rm for} \ i>1
\end{cases}.
\eeq 
The summation is carried out only over the frequency bins in which the
GW signal dominates the noise, $P_g(i)>P_n$, and we define $N_{\rm
  max}$ as the number of the highest frequency bin.

Since we use the discrete power spectra and $N_{\rm max}$ is discrete,
the SNR given in Eq. (\ref{rhopul2}) is a discontinuous function of
model parameters.  This is inevitable as long as the TOA data are
sampled at discrete time intervals.  However, for the Fisher matrix
calculation, we replace the equations by the expression of the
continuous power spectrum, 
\beq
\hat\rho=\sqrt{\frac{M(M-1)/2}{1+\hat{\sigma}^2_{\Delta r}/\hat{\sigma}_g^4\sigma_\zeta^2}},
\label{rhopul3}
\eeq
\beq
\hat{\sigma}_g^2=\frac{\sigma^2_d}{f_h-f_l}\int^{f_{\rm max}}_{f_l}df\frac{P_R(f)}{P_n(f)},
\eeq
\beq
\hat{\sigma}_{\Delta r}^2=\frac{\sigma^4_d}{N(f_h-f_l)}\int^{f_{\rm max}}_{f_l}df\left[1+\left(\frac{P_R(f)}{P_n(f)}\right)^2\bar{\zeta^2}\right].
\eeq
Here, $f_{\rm max}$ is given by $P_R(f_{\rm max})=P_n(f_{\rm max})$.
Then, Eq.  (\ref{rhopul3}) become a smooth function of the parameters,
and is a good approximation of Eq. (\ref{rhopul2}) for $N_{\rm
  max}\gg 1$.  Using Eq. (\ref{rhopul3}), we can calculate the Fisher
matrix as
\beq
\mathcal F_{ij}=\frac{1}{N^2}\frac{\partial S}{\partial \theta_i}\frac{\partial S}{\partial \theta_j},
\label{Fpul}
\eeq
where $S=\left(1+\hat{\sigma}^2_{\Delta
    r}/\hat{\sigma}_g^4\sigma_\zeta^2\right)^{-1/2}$ and
$N=\left(M(M-1)/2\right)^{-1/2}$.  If the interval of the
discontinuity in Eq. (\ref{rhopul2}) is much smaller than the error
width derived from the Fisher matrix Eq. (\ref{Fpul}), it is
consistent to approximate Eq. (\ref{rhopul2}) by Eq. (\ref{rhopul3})
and Eq. (\ref{Fpul}) provides a good prediction.  This is the case in
our fiducial model investigated in the next section.

In this paper, we consider SKA, the future radio telescope array which
is expected to discover a large number of pulsars and observe pulses
with high accuracy, and take the parameters as $N=500, M=100,
\sigma_n=50{\rm ns}, T= 10{\rm years}$, according to
Ref. \cite{Sesana:2010mx}.  The detection threshold is taken as
$\rho>4$ again.  NANOGrav, one of the latest experiments, have placed
the upper bound of the GW background, $\Omega_{\rm GW}<1.9\times
10^{-8}$ for $f\simeq 1/(5{\rm years})=6.3\times10^{-9}{\rm Hz}$,
under the assumption that $\Omega_{\rm GW}$ has a power-law spectrum
\cite{Demorest:2012bv}.

\subsection{Measurement of CMB fluctuation}

\begin{table}[t]
  \begin{center}
  \begin{tabular}{l|c|c|c}
  \hline
  \hline
  bands [GHz] & $\theta_{\rm FWHM}$ [arcmin] & $\sigma_T$ [$\mu$K] & $\sigma_P$ [$\mu$K] \\
  \hline
      $70$ & $14.0$ & $4.7$ & $6.7$ \\
      $100$ & $10.0$ & $2.5$ & $4.0$ \\
      $143$ & $7.1$ & $2.2$ & $4.2$ \\
      $217$ & $5.0$ & $4.8$ & $9.8$ \\
  \hline
  \hline 
\end{tabular}
\caption{Survey parameters adopted in our analysis for Planck.  The
  values are taken from Ref. \cite{Planck:2006aa}.}
  \label{table:planck}
\end{center}
\end{table}

\begin{table}[t]
  \begin{center}
  \begin{tabular}{l|c|c|c}
  \hline
  \hline
  bands [GHz] & $\theta_{\rm FWHM}$ [arcmin] & $\sigma_T$ [$\mu$K] & $\sigma_P$ [$\mu$K] \\
  \hline
      $45$ & $17$ & $5.85$ & $8.27$ \\
      $70$ & $11.0$ & $2.96$ & $4.19$ \\
      $100$ & $8.0$ & $2.29$ & $3.24$ \\
      $150$ & $5$ & $2.21$ & $3.13$ \\
      $220$ & $3.5$ & $3.39$ & $4.79$ \\
  \hline
  \hline 
\end{tabular}
\caption{Survey parameters adopted in our analysis for CMBpol.  The
  values are taken from Ref. \cite{Baumann:2008aq}.  }
  \label{table:CMBpol}
\end{center}
\end{table}

The Fisher matrix for measurement of CMB fluctuation is given
by\cite{Zaldarriaga:1997ch}
\beq
\mathcal F_{ij}=\sum_l\sum_{X,X^{\prime}} \frac{\partial C_l^X}{\partial \theta_i}{\rm Cov}^{-1}(C^X_l,C^{X^{\prime}}_l) \frac{\partial C_l^{X^{\prime}}}{\partial \theta_j},
\eeq
where $X$ and $X^{\prime}$ are summed over the temperature(TT), E-mode
polarization(EE), B-mode polarization(BB) and cross-correlation
between temperature and E-mode(TE).  Cov is the covariance matrix and
given by
\begin{center}
\beq
 {\rm Cov}(C^{TT}_l,C^{TT}_l)=\frac{2}{(2l+1)f_s}(C^{TT}_l+N_{T;l})^2, \nonumber 
 \eeq
 \beq
 {\rm Cov}(C^{EE}_l,C^{EE}_l)=\frac{2}{(2l+1)f_s}(C^{EE}_l+N_{P;l})^2, \nonumber 
  \eeq
  \beq
 {\rm Cov}(C^{BB}_l,C^{BB}_l)=\frac{2}{(2l+1)f_s}(C^{BB}_l+N_{P;l})^2, \nonumber 
  \eeq
  \beq
 {\rm Cov}(C^{TE}_l,C^{TE}_l)=\frac{2}{(2l+1)f_s}\left[(C^{TE}_l)^2+(C^{TT}_l+N_{T;l})(C^{EE}_l+N_{P;l})\right], \nonumber 
 \eeq
  \beq
 {\rm Cov}(C^{TT}_l,C^{EE}_l)=\frac{2}{(2l+1)f_s}(C^{TE})^2, \nonumber 
  \eeq
  \beq
 {\rm Cov}(C^{TT}_l,C^{TE}_l)=\frac{2}{(2l+1)f_s}C^{TE}_l(C^{TT}_l+N_{T;l}), \nonumber
  \eeq
  \beq
 {\rm Cov}(C^{EE}_l,C^{TE}_l)=\frac{2}{(2l+1)f_s}C^{TE}_l(C^{EE}_l+N_{P;l}), \nonumber 
  \eeq
  \beq
  {\rm Cov}(C^{TT}_l,C^{BB}_l)={\rm Cov}(C^{EE}_l,C^{BB}_l)={\rm Cov}(C^{TE}_l,C^{BB}_l) = 0,
   \eeq
\end{center}  
where $f_s$ denotes the sky coverage and is set 0.65 for both Planck
and CMBpol in this paper.  The noise power spectrum $N_{T,P;l}$ is
given by \cite{Knox:1995dq}
\beq
N_{T,P;l}=\left[ \sum_i (N^{(i)}_{T,P;l})^{-1}\right]^{-1}, \ N^{(i)}_{T,P;l}=  \left(\theta^{(i)}_{FWHM}\sigma_{T,P}^{(i)}\right)^2\exp\left(l(l+1)\frac{\left(\theta^{(i)}_{FWHM}\right)^2}{8\ln 2}\right) ,
\eeq
where $\theta^{(i)}_{FWHM}$ is the full width at half maximum of the
Gaussian beam and $\sigma^{(i)}_{T,P}$ is the root mean square of the
instrumental noise per pixel for temperature or polarization, for the
$i$-th frequency band.  The frequency bands and parameter values are
provided in Tables \ref{table:planck} and \ref{table:CMBpol} for
Planck and CMBpol, respectively.

The current limit on $G\mu$ from CMB experiments is derived in
Ref. \cite{Dvorkin:2011aj} using WMAP and SPT data
\cite{Keisler:2011aw}.  Their analysis provides the limit of
$f_{str}<0.0175$, where 
\beq
f_{str}\equiv\frac{\sigma^2_{str}}{\sigma^2_{inf}}, \
\sigma^2_{str}=\sum^{2000}_{l=2}\frac{2l +1}{4\pi}C^{TT,str}_l, \
\sigma^2_{inf}=\sum^{2000}_{l=2}\frac{2l +1}{4\pi}C^{TT,inf}_l.  
\eeq
We find this constraint corresponds in our model to
$G\mu<1.4\times10^{-7}$ for $p=1$, $G\mu<3.6\times 10^{-8}$ for
$p=10^{-1}$ and $G\mu<1.0\times 10^{-8}$ for $p=10^{-2}$.  \footnote{
  Note that the string network model assumed in
  Ref. \cite{Dvorkin:2011aj} is different from that in this paper.
  This causes a small difference in the shape of the CMB spectrum.  We
  neglect this difference and apply their constraint on $f_{tsr}$ to
  our model. }

Future experiments will improve the limit for $G\mu$ by measuring the
B-mode.  The minimum value of $G\mu$ reachable by the B-mode
measurement can be estimated by
$\sqrt{\sigma_{(G\mu)^2}}$\cite{Seljak:2006hi}, where
\beq
  \sigma^{-2}_{(G\mu)^2}=f_s\sum_l\frac{2l+1}{2}(C^{BB,res}_l+N_{P;l})^{-2}\left(\frac{C^{BB,str}_l}{(G\mu)^2}\right)^2,
\eeq
and $C^{BB,res}_l$ is the residual noise of $C^{BB}_l$ after removing
the contamination from the foregrounds.  For the case where the
lensing effect is not removed, $C^{BB,res}_l=C^{BB,len}_l$, we can
probe $G\mu>2.4\times10^{-8}$ for $p=1$, $G\mu>8.0\times10^{-9}$ for
$p=10^{-1}$, and $G\mu>2.6\times10^{-9}$ for $p=10^{-2}$ by Planck,
and $G\mu>1.2\times10^{-8}$ for $p=1$, $G\mu>3.9\times10^{-9}$ for
$p=10^{-1}$, and $G\mu>1.2\times10^{-9}$ for $p=10^{-2}$ by CMBpol.
Even if we consider the case where the lensing effect is perfectly
removed, $C^{BB,res}_l=0$, the above values are not improved
significantly, because the instrumental noise we assume here is larger
than the lensing noise, $C^{BB,len}_l\ll N_{P;l}$.

\section{Constraints on string parameters from future experiments}

In this section, we forecast constraints on the cosmic string
parameters expected from various types of future experiments, using
the Fisher matrix calculations.  We investigate three fiducial models
where different types of experiments complement to determine the
parameters with better accuracy.  Here, we include only string
parameters $G\mu, \alpha, p$ in theoretical parameters $\theta_i$ and
calculate the Fisher matrix for them.  Assuming that the cosmological
parameters are determined with sufficient accuracy, we set them to the
aforementioned values and do not marginalize the likelihood over them
when calculating constraints on the string parameters.  Before that,
we show the accessible parameter space by current and future
experiments.

\subsection{Accessible parameter space of cosmic string search}

\begin{figure}[p]
\begin{minipage}{1\hsize}
\begin{center}
\subfigure[$p=1$]{
\includegraphics[width=70mm]
{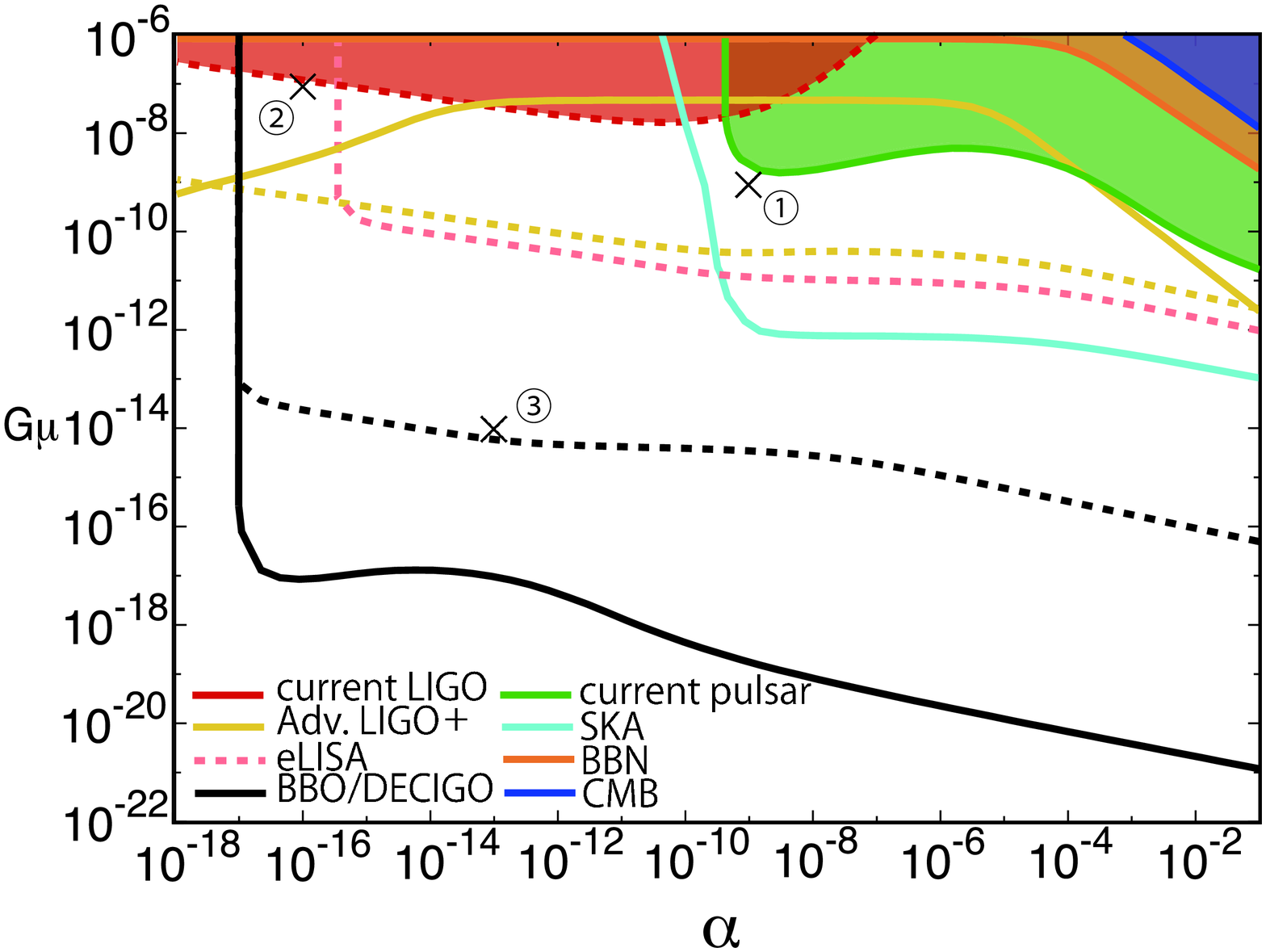}
\label{fig:alpha-Gmu_p=1}
}

\subfigure[$p=10^{-1}$]{
\includegraphics[width=70mm]
{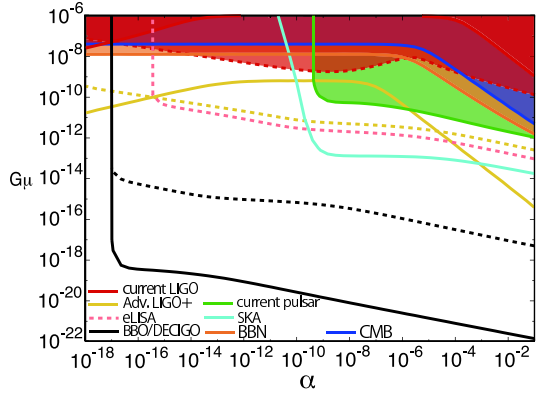}
\label{fig:alpha-Gmu_p=1e-1}
}

\subfigure[$p=10^{-2}$]{
\includegraphics[width=70mm]
{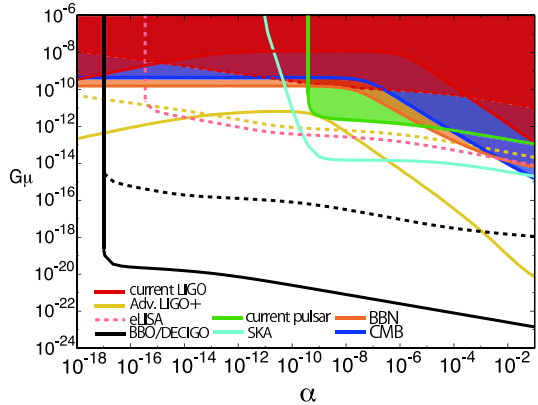}
\label{fig:alpha-Gmu_p=1e-2}
}

\end{center}
\end{minipage}
\caption{Accessible parameter space in the $G\mu$-$\alpha$ plane for
  $p=1$, $p=10^{-1}$ and $p=10^{-2}$.  The colored regions are
  excluded by current experiments or cosmological considerations.  The
  region above the solid or dashed lines can be probed by each future
  experiment.  For GW direct detection experiments, solid lines
  correspond to a background search, and dashed lines correspond to
  burst detection.  Here, ``Adv. LIGO+" means the future
  interferometer network consisting of Advanced LIGO, Advanced Virgo
  and KAGRA and ``current pulsar" means the limit from NANOGrav.  The
  numbered crosses denote the fiducial points studied in
  Secs. \ref{case1}-\ref{case3}.  }
  \label{fig:excluded}
\end{figure}

Figs.  \ref{fig:alpha-Gmu_p=1}, \ref{fig:alpha-Gmu_p=1e-1}, and
\ref{fig:alpha-Gmu_p=1e-2} show the regions in the $\alpha-G\mu$ plane
which are excluded by current experiments or cosmological constraints
and those which can be probed by future experiments for $p=1$,
$10^{-1}$, and $10^{-2}$. 
\footnote{
Similar figures can be found in other papers, such as \cite{Siemens:2006yp}.
Our results in Figure 4 are similar to those in other papers, but ours tend to be somewhat larger, because of the difference in the string network model and the parameters which contains theoretical uncertainty, such as the prefactor of Eq. (\ref{strain}).
}
 We also show the fiducial parameter 
points which we consider in the following Fisher analysis.

For GW burst, the region above each curve represents the parameter
space where bursts from cosmic strings are detectable more than once
per year by each detector.  The threshold amplitude and the most
sensitive frequency $f_{\rm best}$ for each interferometer are
$fh=1.7\times10^{-21}$ and $f_{\rm best}=150{\rm Hz}$ for current
LIGO, $fh=3.4\times10^{-23}$ and $f_{\rm best}=220{\rm Hz}$ for
Advanced LIGO, $fh=5.8\times10^{-21}$ and $f_{\rm
  best}=7.0\times10^{-3}{\rm Hz}$ for eLISA/NGO, and
$fh=1.9\times10^{-24}$ and $f_{\rm best}=0.25{\rm Hz}$ for BBO/DECIGO.
For GW background, the detection threshold is determined whether the
amplitude of the background $\Omega_{\rm GW}$ exceeds the sensitivity
of the detector at $f_{\rm best}$, which is $\Omega_{\rm GW}=7.2\times
10^{-6}$ for current LIGO, $\Omega_{\rm GW}=4.5\times 10^{-9}$ for
Advanced LIGO, and $\Omega_{\rm GW}=9.2\times 10^{-17}$ for
BBO/DECIGO.  We assume that if the lowest frequency of GWs emitted by
strings, $f_{lc}\sim(\alpha t_0)^{-1}$, where $t_0$ is the present age
of the universe, is lower than $f_{\rm best}$, the GWs cannot be
detected either as a background or burst.  The vertical cutoff on the
left side of the curves for eLISA/NGO and BBO/DECIGO correspond to the
value of $\alpha$ which gives $f_{lc}=f_{\rm best}$.  \footnote{ If we
  take into account the fact that each interferometer has sensitivity
  over some frequency range, GWs may be detectable even if
  $f_{lc}>f_{\rm best}$.  This may slightly expand the accessible
  region for large $G\mu$. }

For the latest pulsar timing experiment, we use the constraint on
$\Omega_{\rm GW}$ from NANOGrav.  The parameter space is excluded if
strings predict the GW background larger than $\Omega_{\rm
  GW}=1.9\times 10^{-8}$ at $f=1/(5 {\rm years})$.  Here, we assume
that NANOGrav cannot detect GWs if $f_{lc}>1/(5 {\rm years})$, which
corresponds to the vertical line at the left.  For SKA, we show the
region where SNR exceeds $4$.

We also show the cosmological constraints from CMB and BBN.  The
constraint is derived from the fact that the energy density of the GW
background must be small at the last scattering and BBN, so as not to
distort the fluctuations of the CMB or not to change abundance of
various nuclei.  The CMB constraint is $\int \Omega_{\rm GW}(f) d\ln
f<1.4\times10^{-5}$ at the last scattering \cite{Smith:2006nka} and
the BBN constraint is $\int \Omega_{\rm GW}(f) d\ln f < 1.6\times
10^{-5}$ at the epoch of BBN \cite{Siemens:2006yp,Cyburt:2004yc}. The
lower limit of the integral is determined by the lowest frequency of
the GWs emitted by largest and youngest loops at the time of CMB and
BBN. The upper limit is the frequency of GWs emitted by the earliest
loops, which we assume to be formed at the end of the friction
domination, when the temperature of the Universe is $\sim \sqrt G
\mu$.

From Figs.  \ref{fig:alpha-Gmu_p=1}, \ref{fig:alpha-Gmu_p=1e-1} and
\ref{fig:alpha-Gmu_p=1e-2}, we find that, for large $\alpha$, current
and future pulsar timing experiments are powerful to search for cosmic
strings.  On the other hand, pulsar timing experiments cannot access
to small $\alpha$, and other types of experiments, such as the direct
detection by interferometers or CMB measurement help to set
constraints.  In the following subsections, we investigate future
constraints on the cosmic string parameters from the different types
of experiments by choosing parameter sets indicated in
Fig. \ref{fig:alpha-Gmu_p=1}.

\subsection{Case 1: $G\mu=10^{-9},\alpha=10^{-9},p=1$ --- ground-based interferometers and pulsar timing experiments} \label{case1}

First, we consider the case for $G\mu=10^{-9},\alpha=10^{-9},p=1$.  In
this case, $\alpha$ is large enough for GWs to be detected in the
frequency range of pulsar timing experiments.  For such a value of
$\alpha$, the tension is already severely constrained and
$G\mu=10^{-9}$ is the maximum value allowed by current pulsar timing
constraints.  Because of the small tension, we do not expect detection
of the string signature by future CMB experiments.  However, we
instead expect future pulsar timing experiments such as SKA and the
ground-based interferometers will detect GWs from strings.

\begin{figure}[t]
\begin{tabular}{cc}
\begin{minipage}{0.5\hsize}
\begin{center}
  \subfigure[The burst rate at $f=220{\rm Hz}$.  The bursts in the
  blue region are detectable as rare bursts by the future ground-based
  interferometer network.  ]{
\includegraphics[width=80mm]
{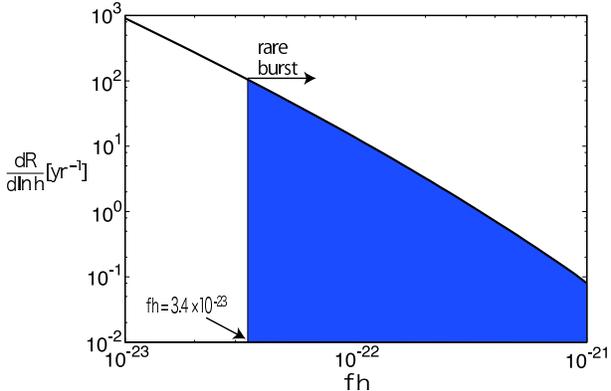}
\label{fig:burst_rate_case1}
}
\end{center}
\end{minipage}

\begin{minipage}{0.5\hsize}
\begin{center}
  \subfigure[The background spectrum $\Omega_{\rm GW}$.]{
\includegraphics[width=80mm]
{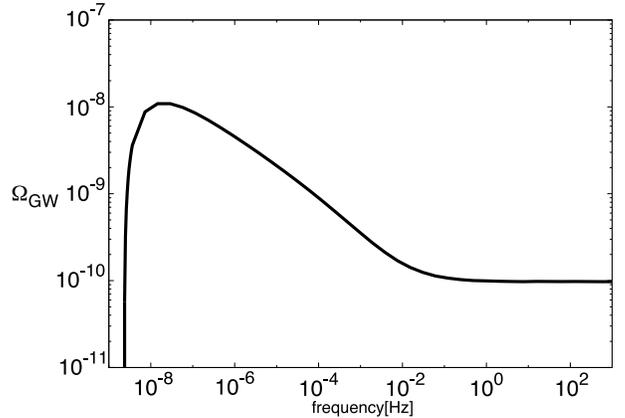}
\label{fig:Omegagw_case1}
}
\end{center}
\end{minipage}
\end{tabular}
\caption{The burst rate at $f=220{\rm Hz}$ and the background spectrum
  $\Omega_{\rm GW}$ for $G\mu=10^{-9}, \alpha=10^{-9}, p=1$.}
\label{fig:case1_spe}
\end{figure}

\begin{figure}[p]
\begin{tabular}{cc}
\begin{minipage}{0.5\hsize}

\subfigure[$G\mu-\alpha$]{
\includegraphics[width=80mm]
{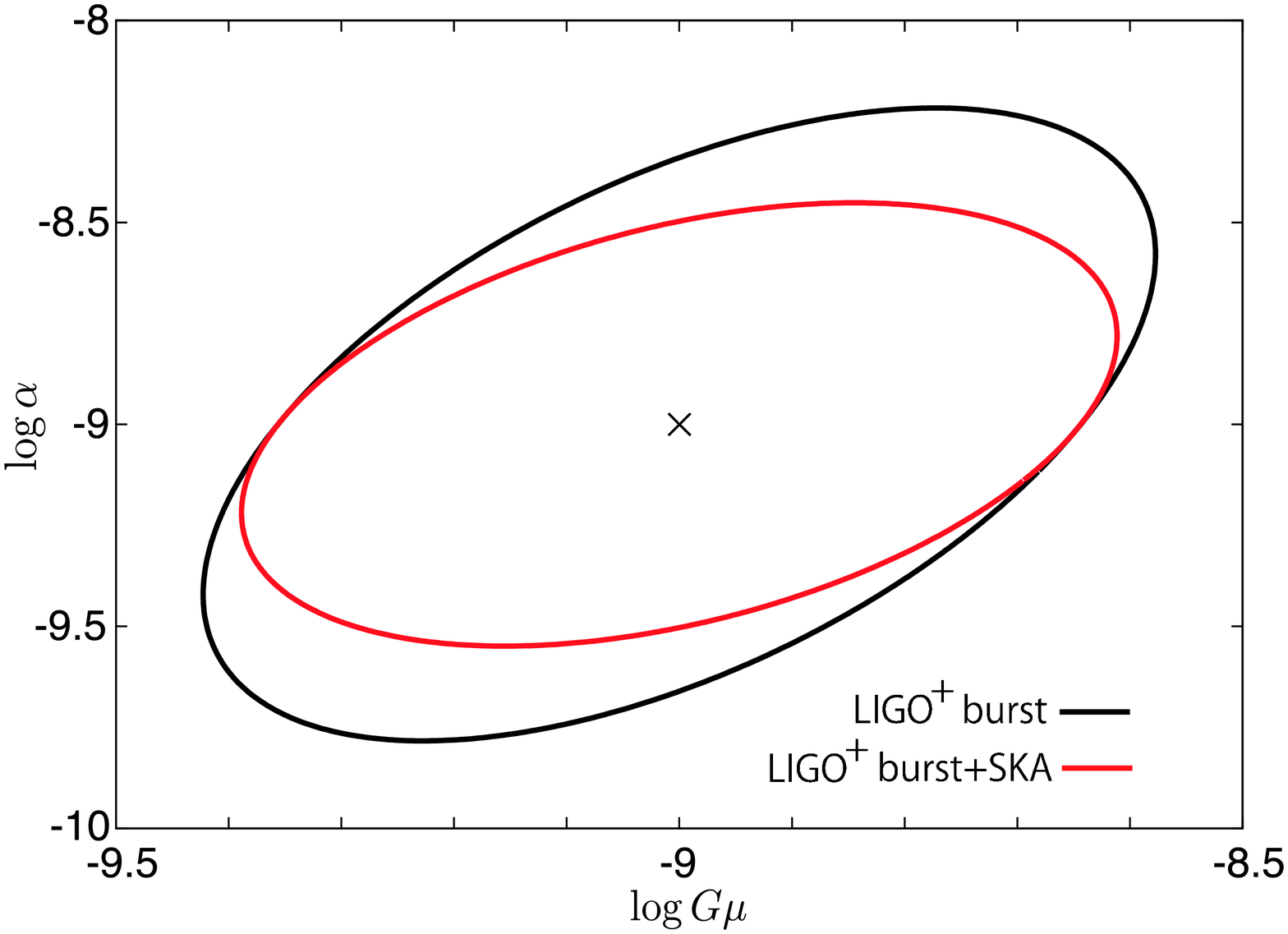}
\label{fig:alpha-Gmu_case1}
}
\end{minipage}

\begin{minipage}{0.5\hsize}
\subfigure[$G\mu-p$]{
\includegraphics[width=80mm]
{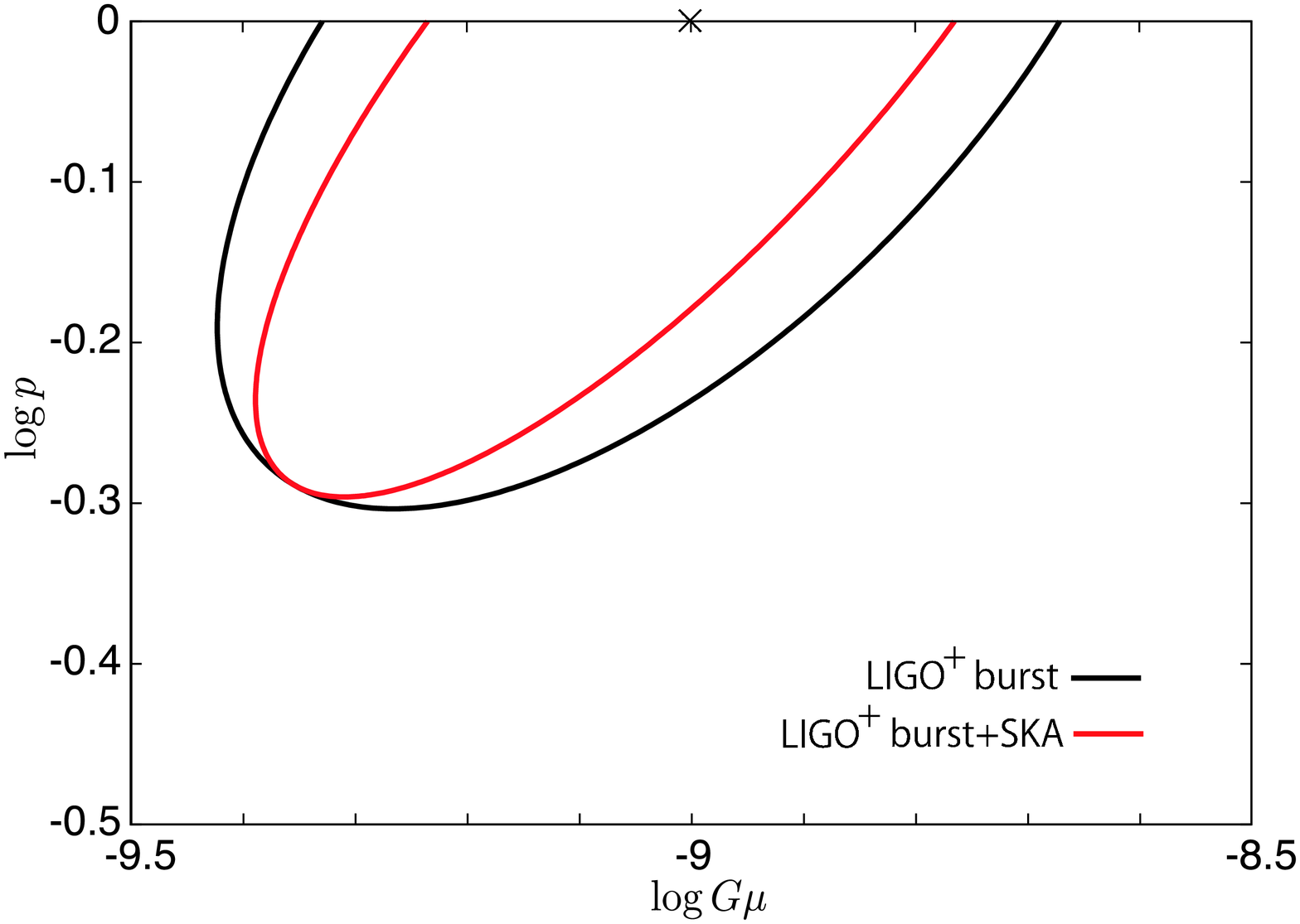}
\label{fig:Gmu-p_case1}
}
\end{minipage}
\\

\begin{minipage}{0.5\hsize}
\subfigure[$\alpha-p$]{
\includegraphics[width=80mm]
{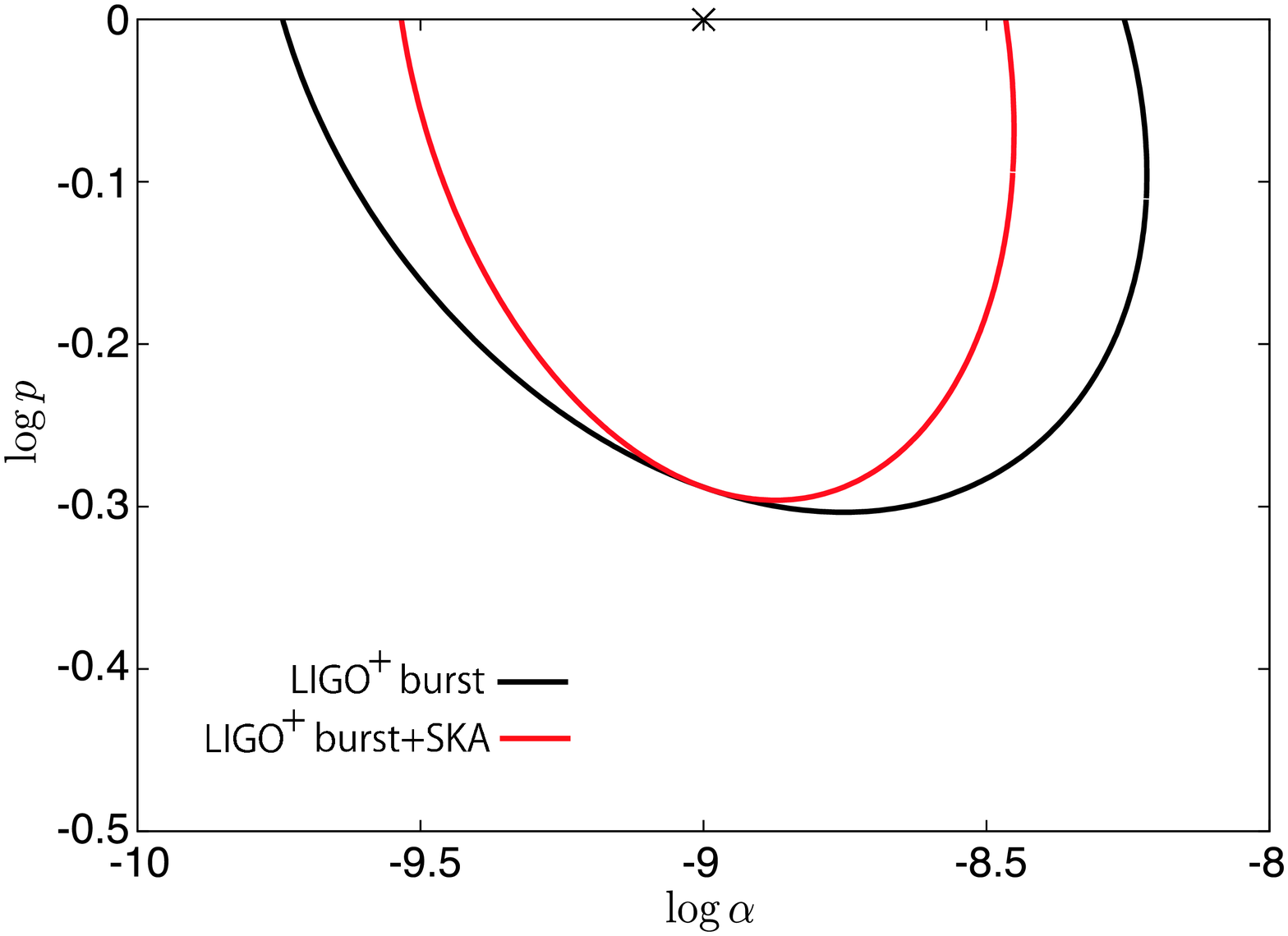}
\label{fig:alpha-p_case1}
}
\end{minipage}
\end{tabular}
\caption{Marginalized $2\sigma$ constraints on cosmic string
  parameters shown in the $G\mu-\alpha$, $G\mu-p$ and $\alpha-p$
  planes.  The fiducial parameter set, denoted by the black cross, is
  taken to be $G\mu=10^{-9}$, $\alpha=10^{-9}$, $p=1$.  The solid
  black line represents the constraints from the burst detection alone
  by the ground-based interferometer network consisting of Advanced
  LIGO, Advanced Virgo and KAGRA.  The red line represents the
  combined constraints from the burst detection and the GW background
  measurement by SKA.}
   \label{fig:case1_cont}
\end{figure}

For this parameter set, future interferometers can detect $168$ rare
bursts with $\rho>4$, where we assume a 3-year run of the
interferometer network consisting of Advanced LIGO, Advanced Virgo and
KAGRA.  However, the GW background is not large enough to be detected
at $f\sim 10^2{\rm Hz}$. In contrast, a 10-year observation by SKA can
detect the GW background with $\rho\sim 33$.  We show the burst rate
$dR/d\ln h$ estimated at $f=220{\rm Hz}$, the most sensitive frequency
of the ground-based interferometers, as a function of burst amplitude
in Fig. \ref{fig:burst_rate_case1}.  We also show the spectrum
$\Omega_{\rm GW}$ in Fig. \ref{fig:Omegagw_case1}.  The plateau region
of the spectrum seen in Fig. \ref{fig:Omegagw_case1} corresponds to
GWs emitted in the radiation-dominated era, and the bump in the low
frequency region corresponds to those emitted after matter-radiation
equality.  Since the energy density of GWs is diluted compared with
the total energy density after matter-radiation equality, the GWs
emitted in the radiation-dominated era is more suppressed compared to
that emitted recently, and this makes background detection at high
frequencies difficult.

In Figs. \ref{fig:alpha-Gmu_case1}, \ref{fig:Gmu-p_case1} and
\ref{fig:alpha-p_case1}, we show the expected constraints on the
string parameters from the burst detection by the ground-based
interferometers and the GW background detection by SKA, estimated by
Fisher matrix calculations.  Since the constraints from SKA alone are
quite weak, we only show the combined constraints in the figure.  One
can see SKA slightly improve the constraints when it is combined to
the constraint from the ground-based interferometers.  This is because
the constraints from pulsar timing and interferometer experiments have
different directions of parameter degeneracy.

Let us briefly discuss the parameter degeneracy.  In this parameter
set, $\alpha < \Gamma G\mu$, so loops evaporate soon after its
formation by emitting GWs of frequency $f\sim (\alpha t)^{-1}$, where
$t$ is the time of the GW emission.  We provide the rough estimates of
the parameter degeneracy in $dR/d\ln h$ and $\Omega_{\rm GW}$ for
$\alpha<\Gamma G\mu$ in Appendix A.  For burst detection, the bursts
whose amplitude is comparable to the sensitivity of the ground-based
interferometers, $fh\sim 3.4\times 10^{-23}$, is in the range of
$h_{3,2}<h<h_{3,3}$ of Eq. (\ref{dR_dh:case3}), where
$fh_{3,2}=2.3\times 10^{-27}$ and $fh_{3,3}=2.2\times 10^{-22}$ in
this case.  So the parameter degeneracy of the burst rate is $\propto
(G\mu)^{3/8}\alpha^{-3/4}p^{-1}$.  Here, $h_{3,2}$ and $h_{3,3}$ are
the corresponding amplitude to characterize when the bursts are
emitted (For details, see Appendix A or
Ref. \cite{Kuroyanagi:2012wm}).  For example, bursts who has amplitude
of $h_{3,2}<h<h_{3,3}$ are emitted between the matter-radiation
equality and $z\simeq 1$.  The rough estimate of the background
spectrum is given by Eq. (\ref{OmegaGW3}).  For the GW background
measurement by SKA, the second term dominates in Eq. (\ref{OmegaGW3}),
so the parameter degeneracy is $\propto
G\mu\alpha^{-1/3}p^{-1}$. \footnote{ Note that, the direction of the
  degeneracy seen in the figures does not directly corresponds to the
  parameter dependence described here, since the shown constraints are
  marginalized over the other parameter.  }.

\begin{figure}[p]
\begin{tabular}{cc}
\begin{minipage}{0.5\hsize}

\subfigure[The burst rate.]{
\includegraphics[width=80mm]
{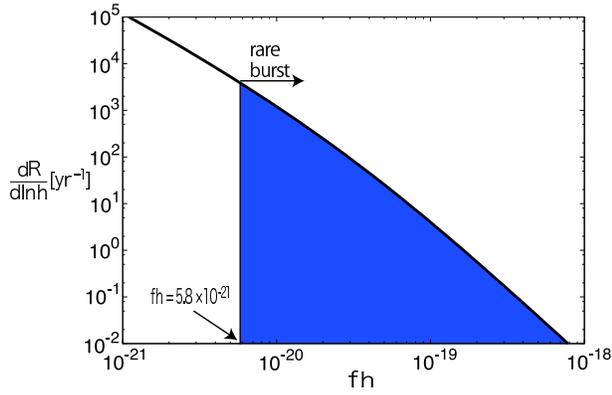}
\label{fig:burst_rate_case1_eLISA}
}
\end{minipage}

\begin{minipage}{0.5\hsize}

\subfigure[$G\mu-\alpha$]{
\includegraphics[width=80mm]
{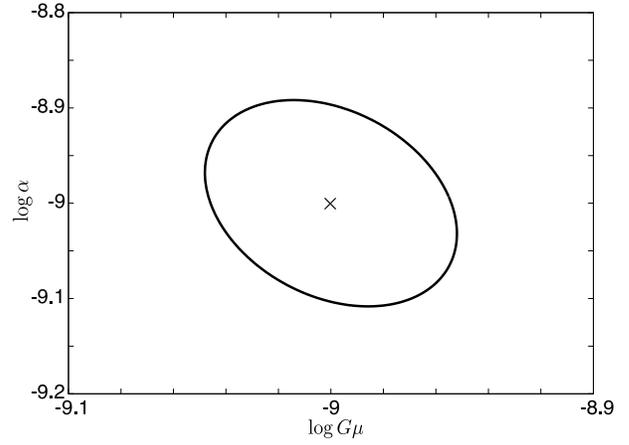}
\label{fig:alpha-Gmu_case1_eLISA}
}
\end{minipage}
\\
\begin{minipage}{0.5\hsize}
\subfigure[$G\mu-p$]{
\includegraphics[width=80mm]
{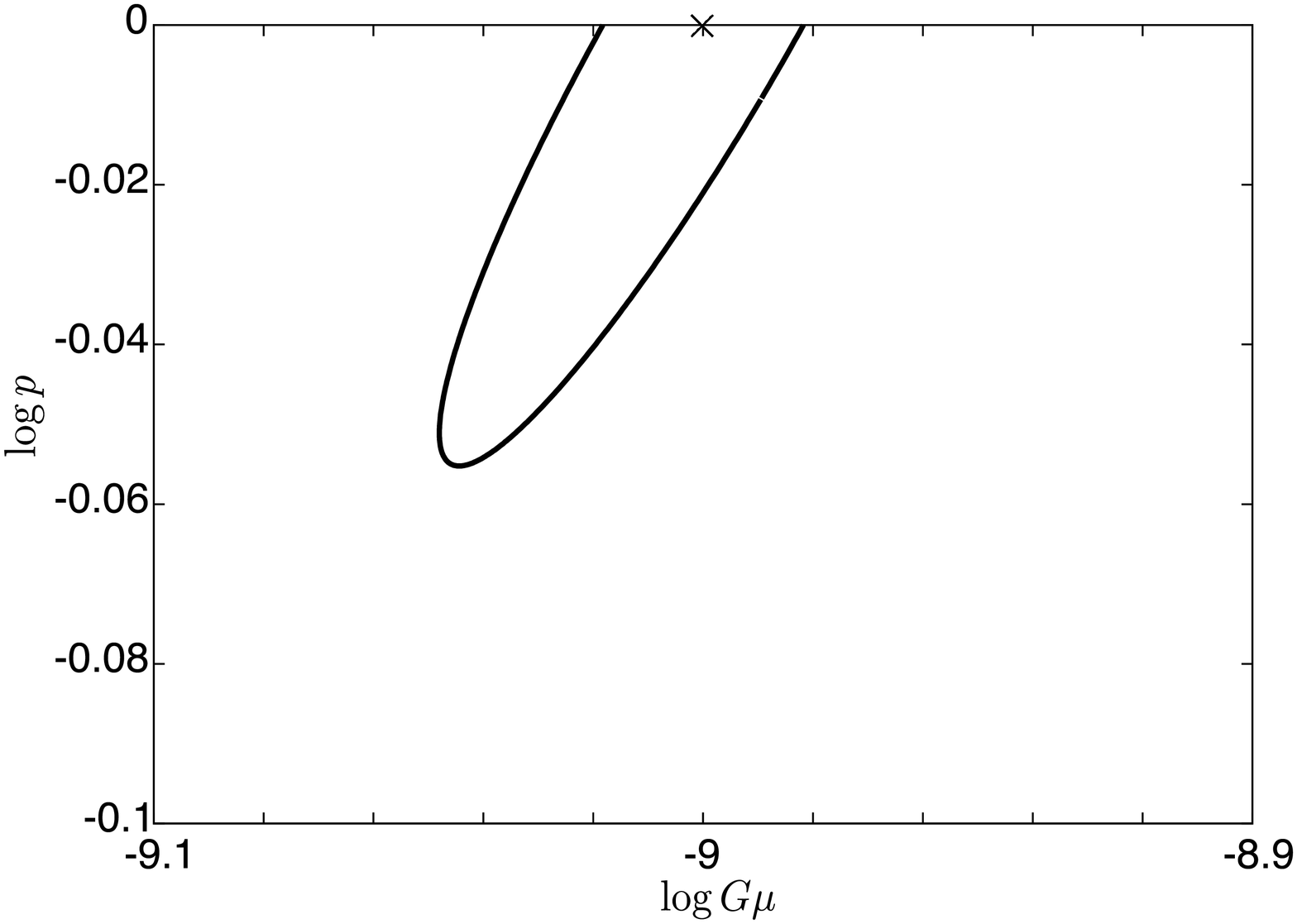}
\label{fig:Gmu-p_case1_eLISA}
}
\end{minipage}

\begin{minipage}{0.5\hsize}
\subfigure[$\alpha-p$]{
\includegraphics[width=80mm]
{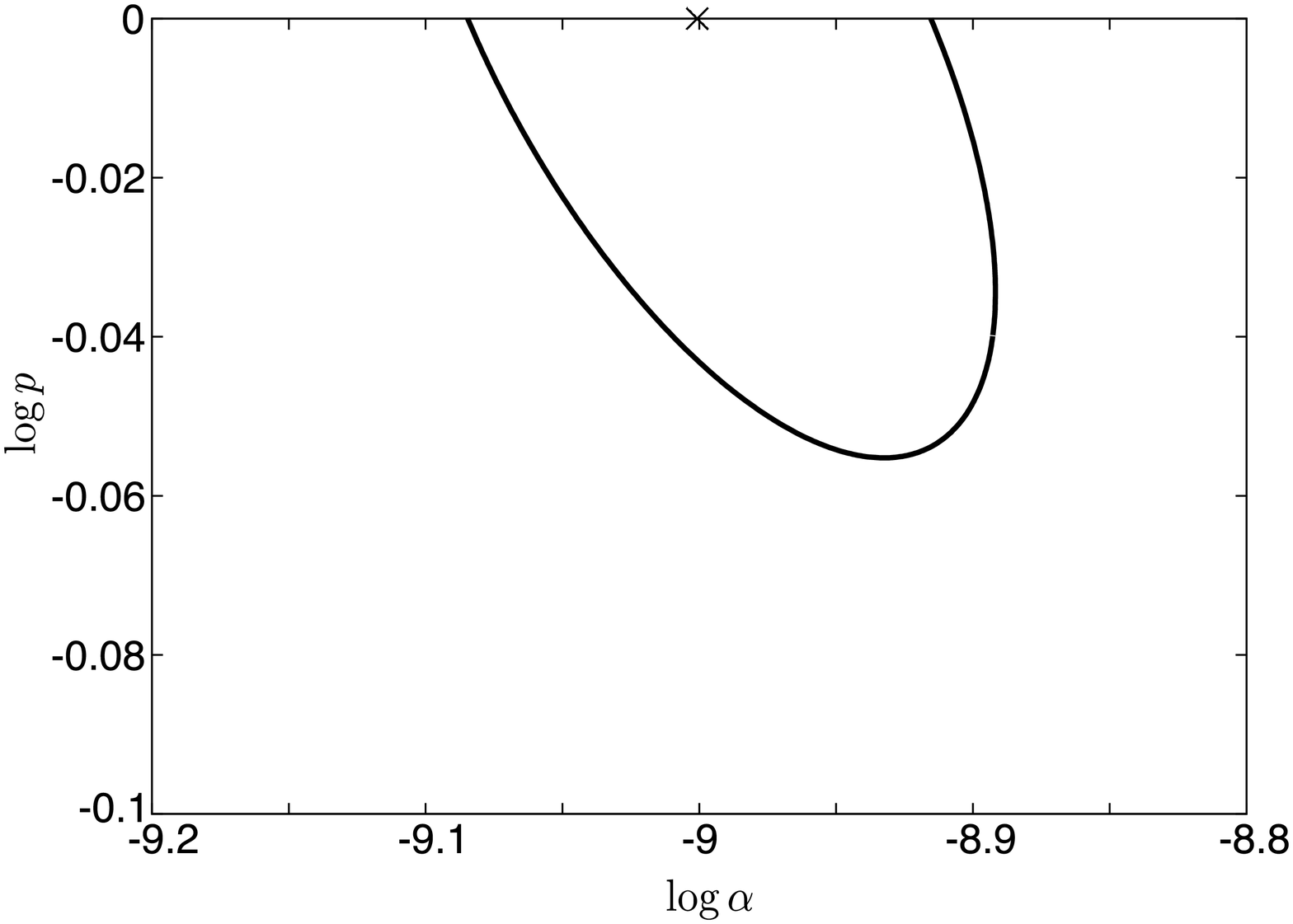}
\label{fig:alpha-p_case1_eLISA}
}
\end{minipage}
\end{tabular}
\caption{(a): The burst rate at $f=7\times10^{-3}{\rm Hz}$ for
  $G\mu=10^{-9}, \alpha=10^{-9}, p=1$.  (b),(c),(d): Marginalized
  $2\sigma$ constraints from eLISA on cosmic string parameters in
  $G\mu-\alpha$, $G\mu-p$ and $\alpha-p$ planes, respectively.  The
  black cross represents the fiducial point.}
\end{figure}

In this fiducial model, eLISA/NGO can also detect GW bursts from
strings.  Fig. \ref{fig:burst_rate_case1_eLISA} shows the burst rate
at the most sensitive frequency, $f=7\times 10^{-3}{\rm Hz}$.  In the
case of eLISA/NGO, the detectable bursts corresponds to the case of
$h>h_{3,3}$ of Eq. (\ref{dR_dh:case3}), where $fh_{3,3}=6.9 \times
10^{-21}$ for $f=7\times 10^{-3}{\rm Hz}$.  Such bursts are emitted
recently at $z\lesssim 1$.  In Figs. \ref{fig:alpha-Gmu_case1_eLISA},
\ref{fig:Gmu-p_case1_eLISA}, and \ref{fig:alpha-p_case1_eLISA}, we
show the constraints expected from a 3-year run of eLISA/NGO, which
makes $1.4\times 10^4$ burst detections with $\rho>4$.  One can
clearly see eLISA/NGO can provide much stronger constraints than
ground-based interferometers and SKA.  And, of course, BBO/DECIGO will
determine the parameters with a significant accuracy.  Our Fisher
calculation indicates the expected errors on the parameters are
$\mathcal{O} (0.1)\%$.
  
\subsection{Case 2: $G\mu=10^{-7},\alpha=10^{-16},p=1$ --- ground-based interferometers and CMB experiments} \label{case2}

\begin{figure}[t]
\begin{tabular}{cc}
\begin{minipage}{0.5\hsize}
\begin{center}
  \subfigure[The burst rate at $f=220{\rm Hz}$.  The bursts in the
  blue region are detectable as rare bursts by the future ground-based
  interferometer network.  The bursts in the orange region form a
  GW background. ]{
\includegraphics[width=80mm]
{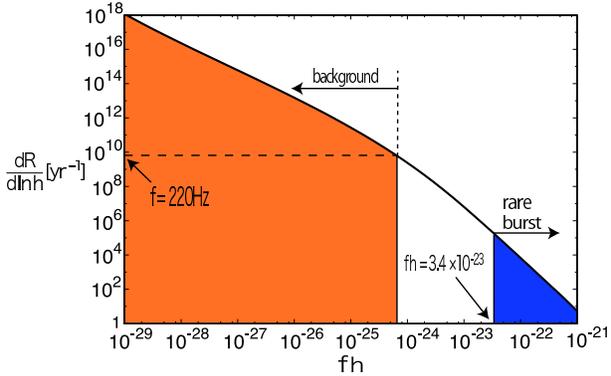}
\label{fig:burst_rate_case2}
}
\end{center}
\end{minipage}

\begin{minipage}{0.5\hsize}
\begin{center}
  \subfigure[The background spectrum $\Omega_{\rm GW}$.]{
\includegraphics[width=80mm]
{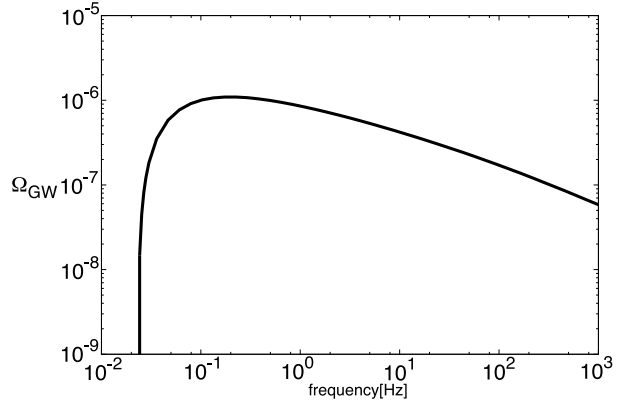}
\label{fig:Omegagw_case2}
}
\end{center}
\end{minipage}
\end{tabular}
\caption{Same as Fig. \ref{fig:case1_spe} but for $G\mu=10^{-7},
  \alpha=10^{-16}, p=1$.}
\end{figure}

Next, we study the case where $G\mu=10^{-7},\alpha=10^{-16},p=1$.  In
this case, $\alpha$ is extremely small, so pulsar timing experiments
cannot detect GWs from strings.  This means that the tension is not
constrained strongly by the current pulsar timing experiments and, if
$G\mu\sim 10^{-7}$ which is still allowed by current CMB and LIGO
experiments, we can expect future CMB experiments to find string
signatures.  Also, we can expect both burst and background detection
by future ground-based interferometers.  The ground-based
interferometers will detect $1.8\times 10^5$ rare bursts with $\rho>4$
and the GW background with $\rho\simeq 187$, where we again assume a
3-year run of the world-wide interferometer network.  We show the
burst rate at $f=220{\rm Hz}$ in Fig. \ref{fig:burst_rate_case2} and
the background spectrum $\Omega_{\rm GW}$ in Fig
\ref{fig:Omegagw_case2}.

\begin{figure}[p]
\begin{tabular}{cc}
\begin{minipage}{0.5\hsize}

\subfigure[$G\mu-\alpha$]{
\includegraphics[width=80mm]
{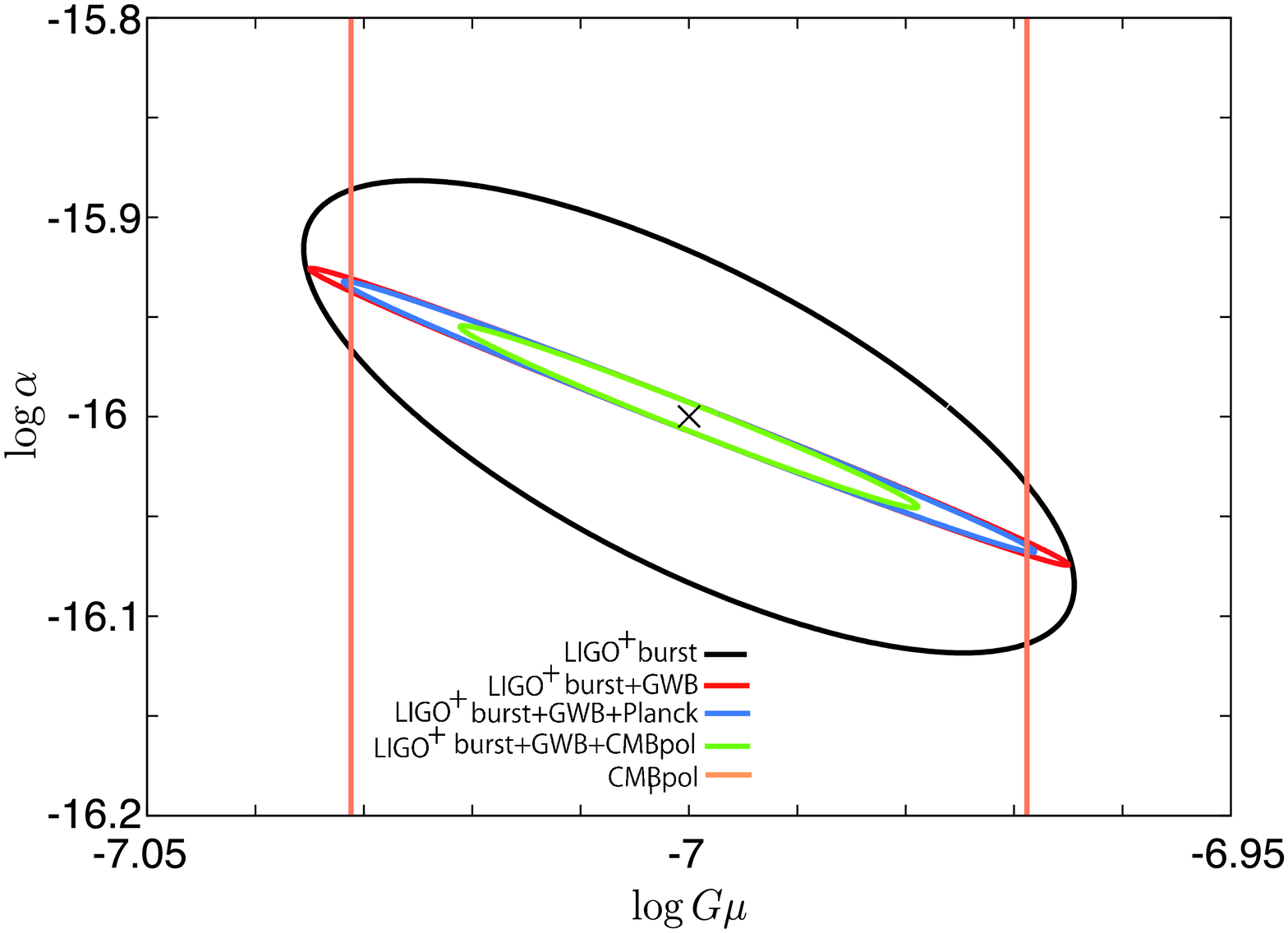}
\label{fig:alpha-Gmu_case2}
}
\end{minipage}

\begin{minipage}{0.5\hsize}
\subfigure[$G\mu-p$]{
\includegraphics[width=80mm]
{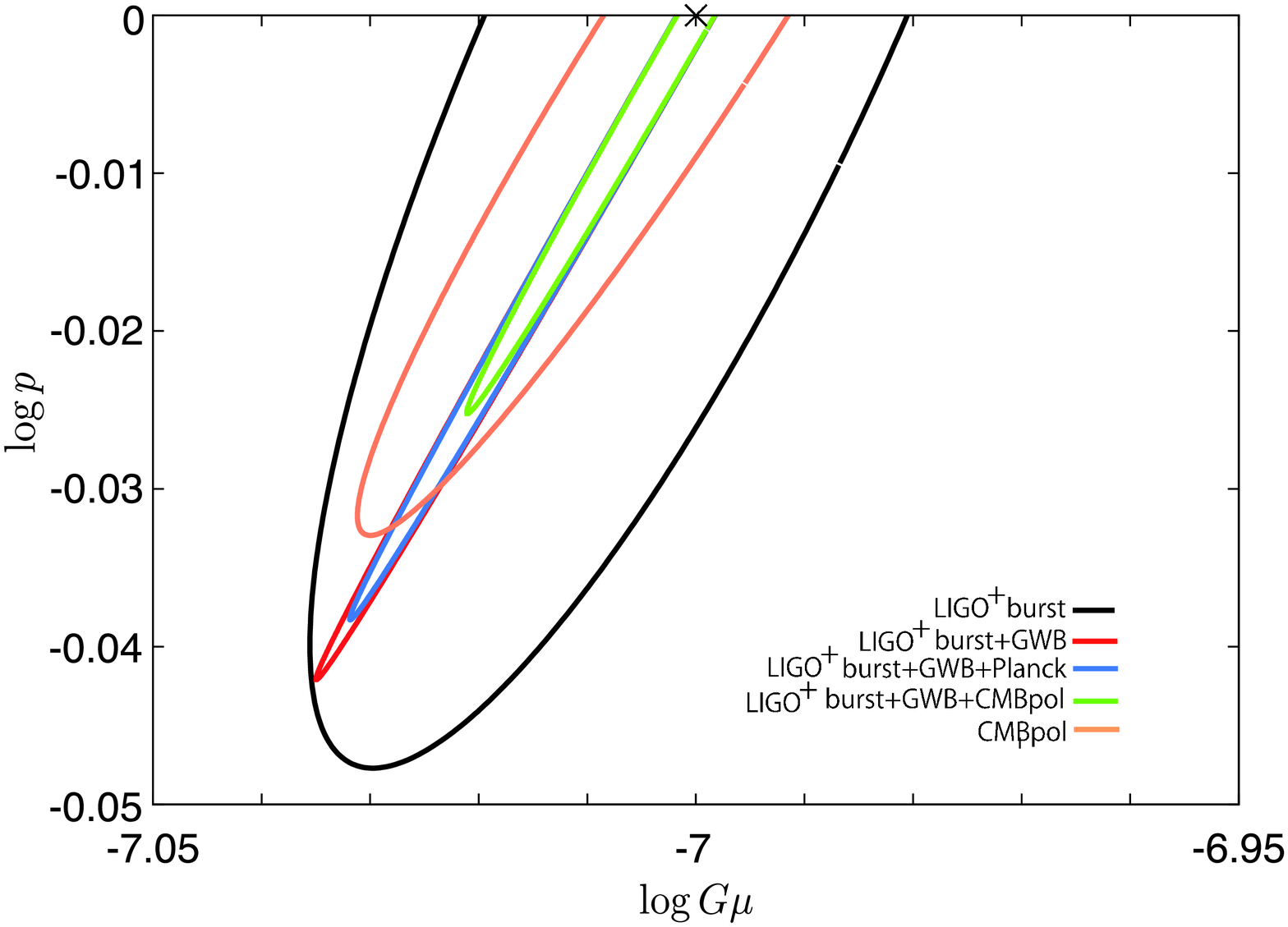}
\label{fig:Gmu-p_case2}
}
\end{minipage}
\\

\begin{minipage}{0.5\hsize}
\subfigure[$\alpha-p$]{
\includegraphics[width=80mm]
{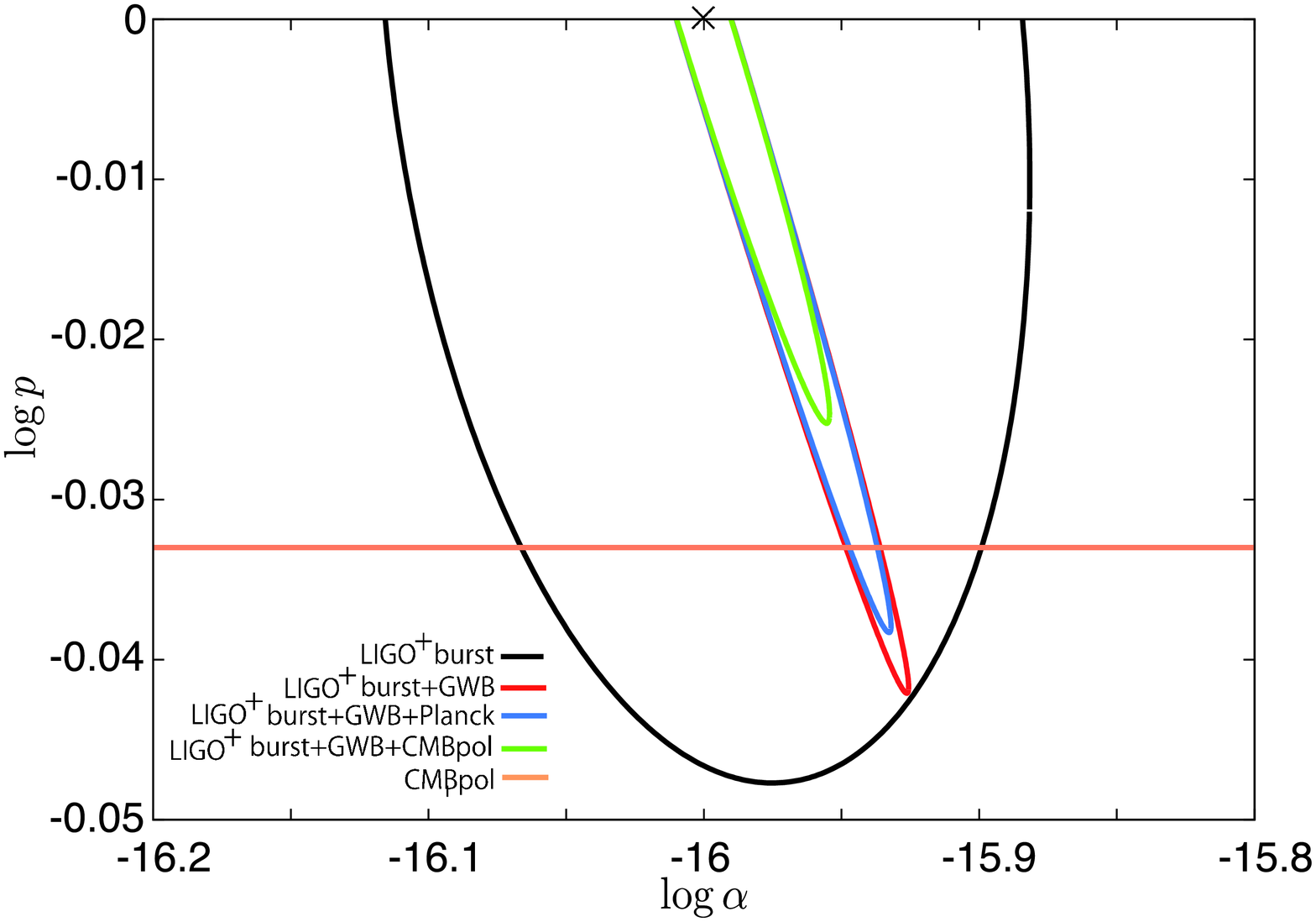}
\label{fig:alpha-p_case2}
}
\end{minipage}
\end{tabular}
\caption{Marginalized $2\sigma$ constraints on cosmic string
  parameters shown in the $G\mu-\alpha$, $G\mu-p$ and $\alpha-p$
  planes.  The fiducial parameter set, denoted by the black cross, is
  taken to be $G\mu=10^{-7}$, $\alpha=10^{-16}$, $p=1$. The solid
  black line represents the constraints from the burst detection by
  the interferometer network consisting of Advanced LIGO, Advanced
  Virgo and KAGRA alone.  The red line represents the combined
  constraints from the burst detection and the GW background
  measurement by the interferometer network.  The blue or green lines
  are the constraints from combination of interferometers and Planck
  or CMBpol, respectively.  The orange lines are the constraint from
  CMBpol alone.  }
\end{figure}

In Figs. \ref{fig:alpha-Gmu_case2}, \ref{fig:Gmu-p_case2} and
\ref{fig:alpha-p_case2}, we show the constraints on the string
parameters from the CMB observation and the burst and GW background
detection by the ground-based interferometers.  For the GW background,
we use the weak signal approximation, Eq. (\ref{FGWBweak}), to
calculate the Fisher matrix.  For CMB, we consider the constraints
from Planck and CMBpol.  We derive the constraints from CMBpol
neglecting the lensing effect.  The results are not significantly
affected by including the lensing effect, as discussed in Sec. 3.4.

In this case, loops are extremely short-lived.  The bursts whose
amplitude is comparable to the sensitivity of the ground-based
interferometers, $fh\sim 3.4\times 10^{-23}$, corresponds to
$h>h_{3,3}$ in Eq. (\ref{dR_dh:case3}), where $fh_{3,3}=4.7\times
10^{-25}$ in this case.  These bursts are emitted recently at
$z\lesssim 1$.  Therefore, for the burst detection, the direction of
the parameter degeneracy is $\propto (G\mu)^{2}\alpha^{1/3}p^{-1}$.
The background spectrum is again expressed by Eq. (\ref{OmegaGW3}).
The bump-like spectrum in Fig. \ref{fig:Omegagw_case2} corresponds to
the second term in Eq. (\ref{OmegaGW3}), which represents GWs emitted
recently, and its parameter dependence is $\propto
G\mu\alpha^{-1/3}p^{-1}$.  CMB measurements provide information on
infinite strings, which is characterized by only $G\mu$ and $p$, and
do not contain information on $\alpha$.  The overall amplitude of CMB
spectra is proportional to $(G\mu)^2$ and decreasing $p$ leads to
enhancement of the amplitude and the change of the spectral shape as
explained in Sec. 3.  We numerically find that the dominant
contribution to the Fisher matrix comes from the temperature spectrum
around $1000\lesssim l \lesssim 2000$.  For such values of $l$,
$C^{TT,str}_l$ is roughly proportional to $p^{-2}$, as shown in
Fig. \ref{fig:CT_p}.  Therefore, the parameter degeneracy of the CMB
constraint is $\propto (G\mu)^2p^{-2}$.

\subsection{Case 3: $G\mu=10^{-14},\alpha=10^{-13},p=1$ ---
  BBO/DECIGO} \label{case3}

\begin{figure}[t]
\begin{tabular}{cc}
\begin{minipage}{0.5\hsize}
\begin{center}
  \subfigure[The burst rate at $f=0.25{\rm Hz}$.  The bursts in the
  blue region are detectable as rare bursts by the future ground-based
  interferometer network.  The bursts in the orange region form a
  GW background. ]{
\includegraphics[width=80mm]
{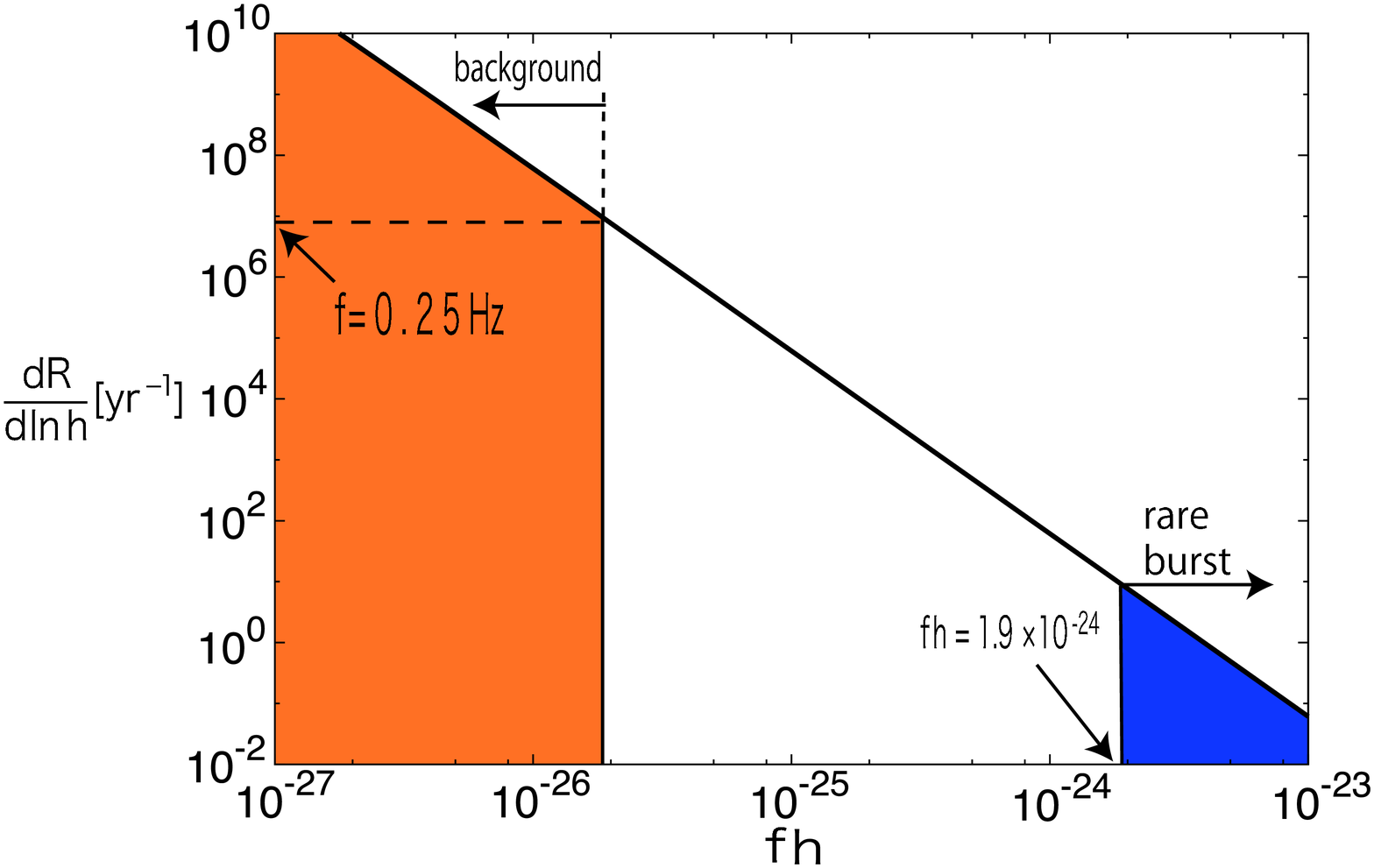}
\label{fig:burst_rate_case3}
}
\end{center}
\end{minipage}

\begin{minipage}{0.5\hsize}
\begin{center}
  \subfigure[The background spectrum $\Omega_{\rm GW}$.]{
\includegraphics[width=80mm]
{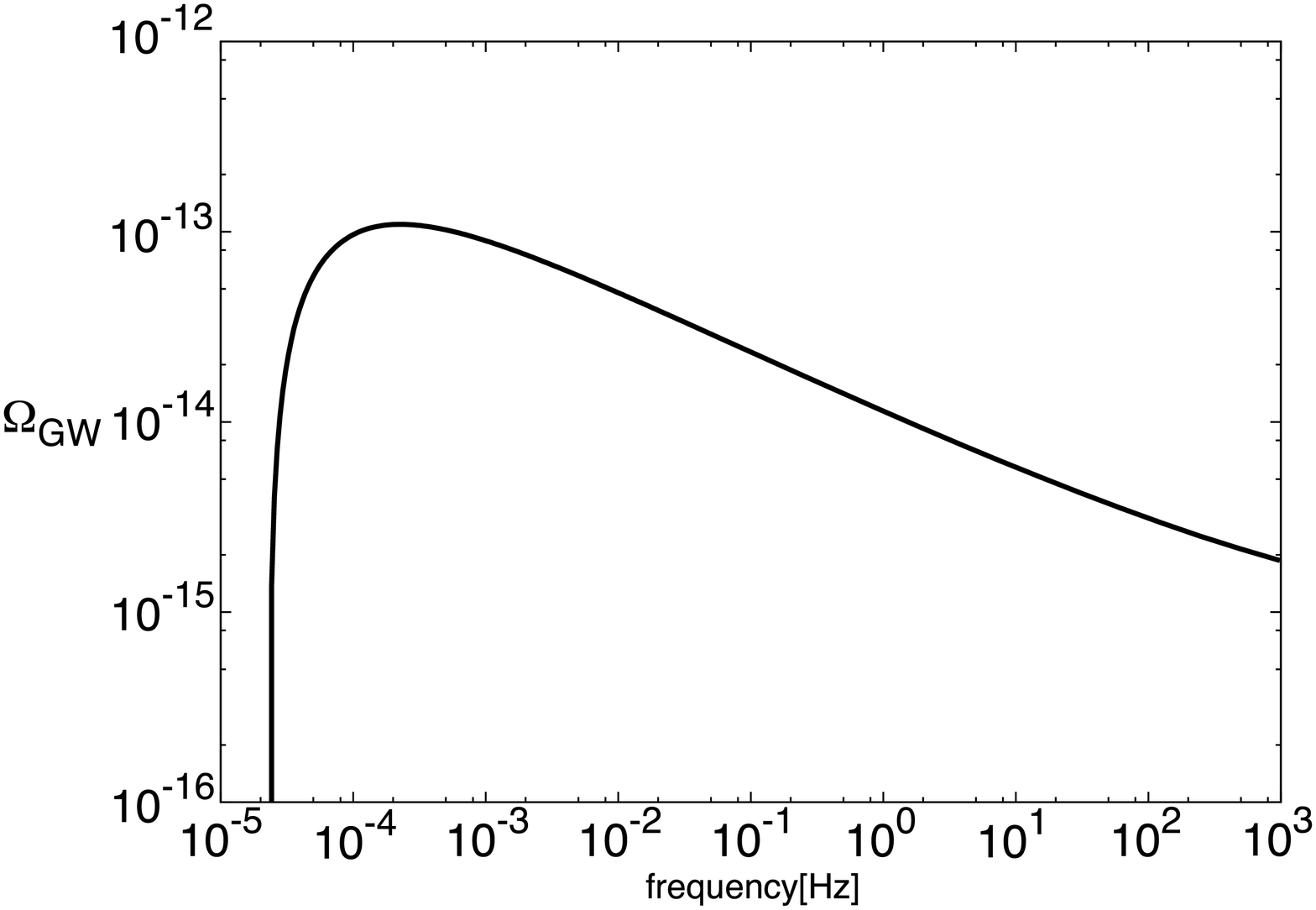}
\label{fig:Omegagw_case3}
}
\end{center}
\end{minipage}
\end{tabular}
\caption{Same as Fig. \ref{fig:case1_spe} but for $G\mu=10^{-14},
  \alpha=10^{-13}, p=1$.}
\end{figure}

Finally, we study the case where $G\mu=10^{-14},\alpha=10^{-13},p=1$.
For this small value of $G\mu$, only BBO/DECIGO can detect string
signals.  In this case, BBO/DECIGO detects $35$ rare bursts with
$\rho>4$ in a 3-year run and measure the GW background with very high
SNR, $\rho\simeq 510$.  We show the burst rate estimated at the best
frequency of BBO/DECIGO, $f=0.25{\rm Hz}$, in
Fig. \ref{fig:burst_rate_case3} and the background spectrum in
Fig. \ref{fig:Omegagw_case3}.

\begin{figure}[p]
\begin{tabular}{cc}
\begin{minipage}{0.5\hsize}

\subfigure[$G\mu-\alpha$]{
\includegraphics[width=80mm]
{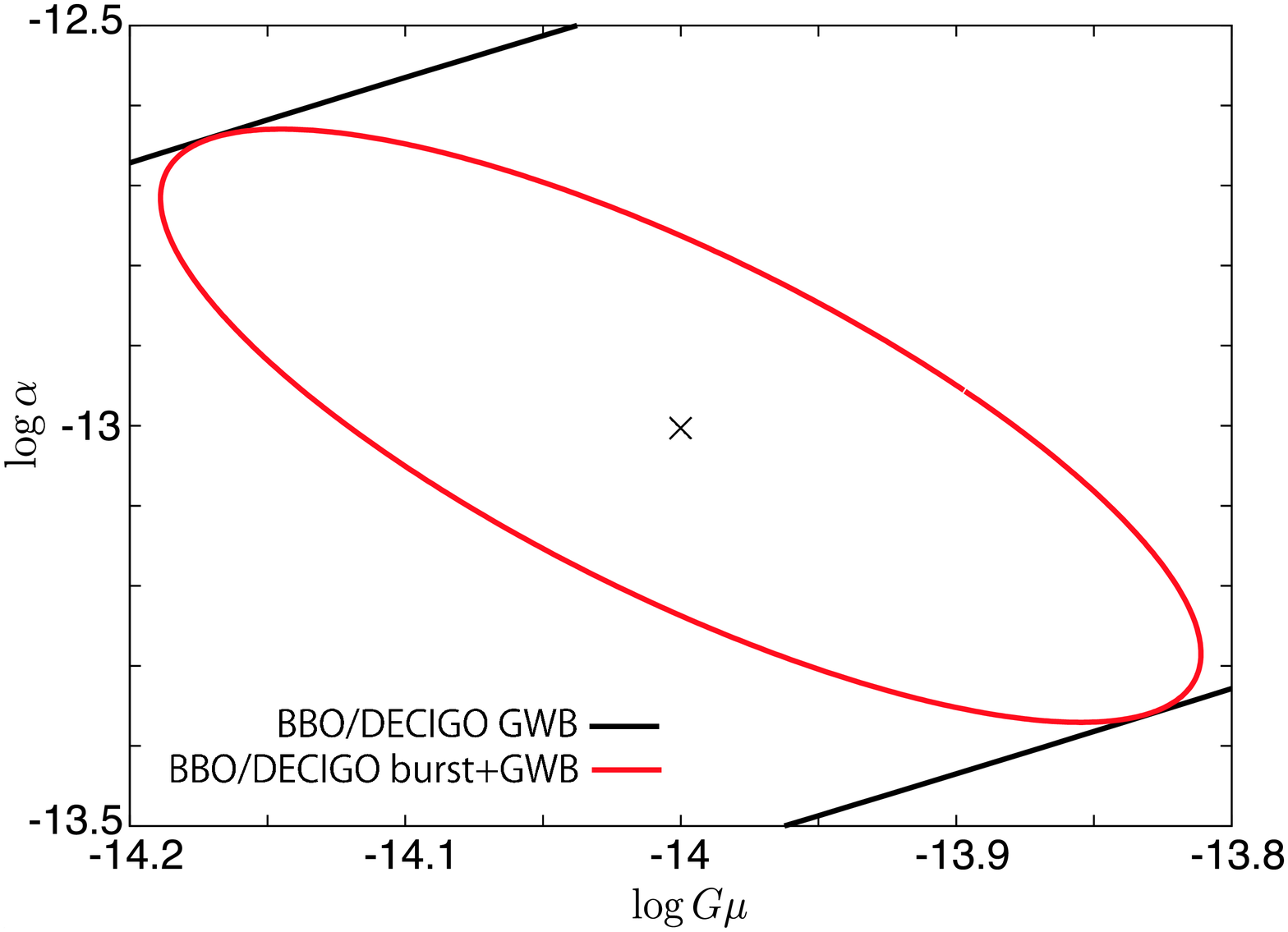}
\label{fig:alpha-Gmu_case3_BBO}
}
\end{minipage}

\begin{minipage}{0.5\hsize}
\subfigure[$G\mu-p$]{
\includegraphics[width=80mm]
{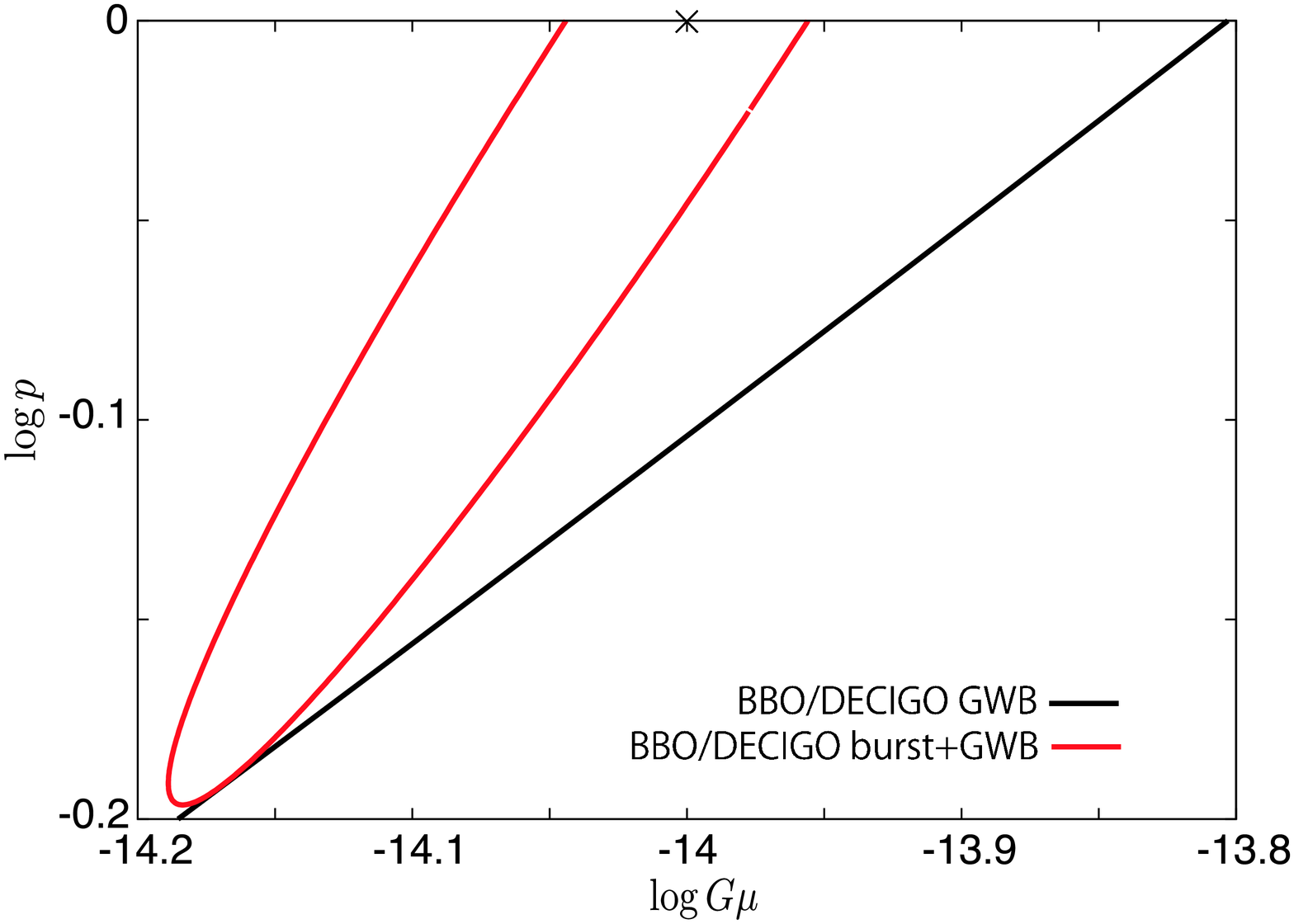}
\label{fig:Gmu-p_case3_BBO}
}
\end{minipage}
\\

\begin{minipage}{0.5\hsize}
\subfigure[$\alpha-p$]{
\includegraphics[width=80mm]
{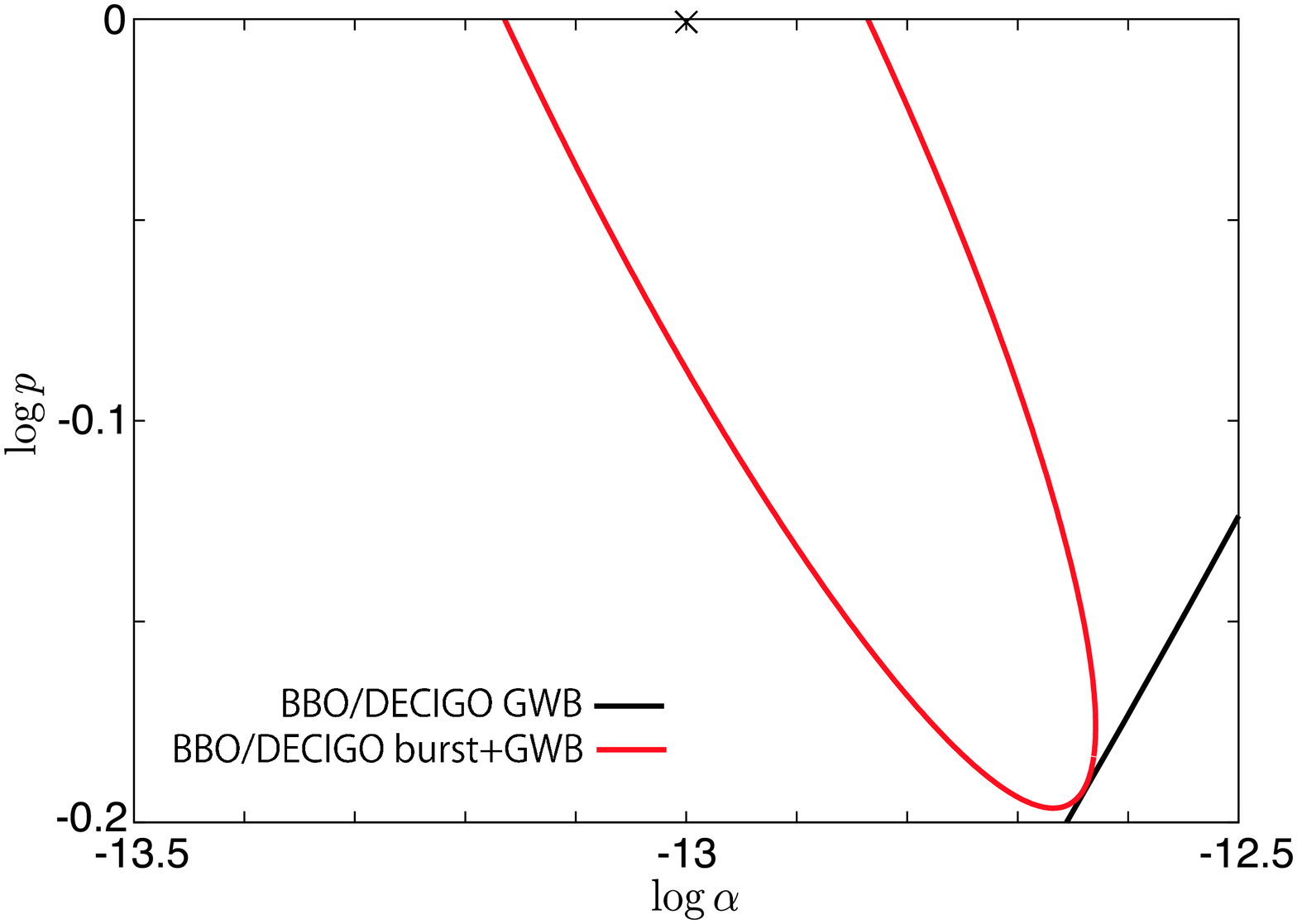}
\label{fig:alpha-p_case3_BBO}
}
\end{minipage}
\end{tabular}
\caption{ The $2\sigma$ constraints from BBO/DECIGO on cosmic string
  parameters in the $G\mu-\alpha$, $G\mu-p$ and $\alpha-p$ planes.
  The black line shows the marginalized $2\sigma$ constraints from
  background measurement alone.  The red line represents the combined
  constraints from burst detection and background measurement.  The
  fiducial parameter set, denoted by the black cross, is taken to be
  $G\mu=10^{-14}$, $\alpha=10^{-13}$, $p=1$.}
   \label{fig:BBOconstraint}
\end{figure}

Figures \ref{fig:alpha-Gmu_case3_BBO}, \ref{fig:Gmu-p_case3_BBO} and
\ref{fig:alpha-p_case3_BBO} show the expected constraints on the
string parameters from the burst detection and the background
measurement by a 3-year run of BBO/DECIGO.  Here, we use the exact
formula of the Fisher matrix for the GW background,
Eq. (\ref{FGWBexact}), because of the high SNR.  The background
measurement provides stronger constraints than burst detection.
However, the constraints from background measurement intrinsically
have strong parameter degeneracies, since the information is
practically only one $\Omega_{GW}$ at $f=f_{\rm best}$.  Thus,
although constraints from the burst detection is weak because of the
small number of detectable events, it dramatically tightens the errors
when combined with the constraints from the background detection.
This is again thanks to the difference in the parametric dependencies.

In this parameter set, loops are marginally short-lived.  The bursts
detectable by BBO/DECIGO are emitted recently, $z< 1$, and their rate
is expressed by the case of $h>h_{3,3}$ of Eq. (\ref{dR_dh:case3}),
where $fh_{3,3}=4.5\times 10^{-28}$.  So the parameter degeneracy is
$\propto (G\mu)^2\alpha^{1/3}p^{-1}$.  The background spectrum around
$f=0.25{\rm Hz}$ corresponds to GWs emitted at $z< 1$ and is roughly
expressed by the second term of Eq. (\ref{OmegaGW3}), whose parameter
dependence is $\propto G\mu \alpha^{-1/3} p^{-1}$.

\section{Summary}

Among many types of cosmic string signatures, gravitational waves from
cosmic string loops and CMB fluctuations induced by infinite strings
are important and future experiments will help to test the existence
of cosmic strings.  If detected, we may even able to extract
information on the nature of cosmic strings.  In this paper, we extend
our previous work which investigated the constraints on the string
parameters from future ground-based GW experiments.  In addition to
the ground-based interferometers, this paper have investigated
constraints from space-borne interferometers, pulsar timing arrays and
CMB experiments.  Furthermore, we have studied the combination of
information from these observations.

Each experiment sheds light on the different aspect of cosmic strings
and gives us different information on strings.  More technically
speaking, the constraints from different experiments have different
parameter degeneracies.  CMB experiments probe infinite strings
through their gravitational effects on the background photons, while
GW interferometers and pulsar timing experiments probe string loops by
detecting GWs emitted from them.  Among GW experiments, each of them
targets a different frequency band; ground-based interferometers
detect GWs of $f\sim 220{\rm Hz}$, space-borne ones probe those of
$f\sim 10^{-4}$ to $0.1{\rm Hz}$ and pulsar timing experiments are
sensitive to those of $f\sim 10^{-8}{\rm Hz}$.  GWs of different
frequency are emitted at different epoch of the Universe and provide
us with independent information.  Besides, different types of GW
observations, the burst detection and the background measurement,
provide different information on cosmic strings.  These are the
reasons why we can break the parameter degeneracies by combining these
experiments and obtain better constraints on cosmic string parameters.

In this paper, we have studied three different fiducial models, where
different types of experiments are help each other to constrain the
string parameters.  The first case is
$G\mu=10^{-9},\alpha=10^{-9},p=1$, where both future GW
interferometers and pulsar timing such as SKA can detect GWs from
cosmic strings.  We also calculate constraints from eLISA/NGO and
found that it is more efficient to constrain string parameters than
the pulsar timing and ground-based experiments.  The second case is
$G\mu=10^{-7},\alpha=10^{-16},p=1$, where both future GW
interferometers and CMB experiments can detect string signatures.  The
third case is $G\mu=10^{-14},\alpha=10^{-13},p=1$, where the tension
is so small that only ultimate space-borne interferometers such as BBO
and DECIGO can detect string signatures.  We have shown that future GW
interferometers, especially space-borne ones, are very powerful to
investigate cosmic strings not only because of their extreme
sensitivity but also because they can probe strings in two different
ways, the background measurement and the burst detection.

Finally, we should note that, in the case where we can determine the
parameters with a very good accuracy, uncertainties in the string
network model become more important.  In that case, further
theoretical study will be needed to perform analysis with more
accurate modeling, or precise measurements of cosmic string GWs by
future space-borne interferometers may even be able to shed light on
the theoretical uncertainties in the string evolution.

\appendix

\section*{Appendix}

\section{ Parameter dependence of the burst rate and the GW background
  spectrum}
\label{app1}  
In this appendix, we provide rough estimates of the burst rate
$dR/d\ln h$ and the GW background spectrum $\Omega_{\rm GW}$ to
understand the dependence of these quantities on the cosmic string
parameters.  We concentrate on the case of $\alpha<\Gamma G\mu$, where
loops are short-lived, since this case applies to all the fiducial
models investigated in Sec. 4.  For other cases, see the appendices of
Ref. \cite{Kuroyanagi:2012wm}.

In this case, $\alpha<\Gamma G\mu$, the burst rate is given by
\small
\beq
\frac{dR}{d\ln h} (f,h) \sim
\begin{cases}
&  (G\mu)^{6/5}\alpha^{-1/5} \gamma_r^{-2} \left (\frac{\Omega_m}{\Omega_r}\right)^{-11/10} f^{-18/5} t_0^{-12/5} h^{-11/5}\\
& \qquad  ;G\mu\alpha^{-1}f^{-3}t_0^{-2}\left (\frac{\Omega_m}{\Omega_r}\right)^{-1/2}\equiv h_{3,1}<h<G\mu\alpha^{2/3} f^{-4/3} t_0^{-1/3} \left (\frac{\Omega_m}{\Omega_r}\right)^{-4/3} \equiv h_{3,2} \\
&  (G\mu)^{3/8}\alpha^{-3/4} \gamma_m^{-2} f^{-5/2} t_0^{-17/8} h^{-11/8} \\
& \qquad  ;h_{3,2}<h<G\mu\alpha^{2/3}f^{-4/3}t_0^{-1/3}\equiv h_{3,3} \\
&  (G\mu)^{2}\alpha^{1/3} \gamma_m^{-2} f^{-14/3} t_0^{-8/3} h^{-3}  \qquad ;h>h_{3,3}
\end{cases}.
\label{dR_dh:case3}
\eeq
\normalsize
Bursts in the range of $h_{3,1}<h<h_{3,2}$, $h_{3,2}<h<h_{3,3}$, and
$h>h_{3,3}$ are emitted in the radiation-dominated era, in the
matter-dominated era, and at $z\ll 1$, respectively.  The burst rate
for $h<h_{3,1}$ is suppressed strongly.

The background spectrum is given by 
\beq
\Omega_{\rm GW}(f)\sim G\mu \gamma_r^{-2} \frac{\Omega_r}{\Omega_m} + G\mu\gamma_m^{-2}\alpha^{-1/3}t_0^{-1/3}f^{-1/3}. \label{OmegaGW3}
\eeq
Here, the first term is the contribution from GWs emitted in the
radiation-dominated era and the second one represents GWs emitted
recently.  Note that the former has smaller amplitude than the latter
by the factor of $\Omega_r/\Omega_m$, since the energy density of GWs
decays faster than the total energy density in the matter-dominated
era.

\section*{Acknowledgment}

K.M. and T.S. would like to thank the Japan Society for the Promotion
of Science for financial support.  This work is supported by the
Grant-in-Aid for Scientific Research from the Ministry of Education,
Science, Sports, and Culture (MEXT), Japan, No. 23340058 (S.K.),
No. 24740149 (S.K.), No.~23740179 (K.T.), No.~24111710 (K.T.) and
No.~24340048 (K.T.).

{}

\end{document}